\documentclass[conference]{IEEEtran}
\IEEEoverridecommandlockouts
\usepackage{amsthm}
\usepackage{amsfonts,amsmath}
\usepackage{algorithm}
\usepackage[noend]{algpseudocode}
\usepackage{bm}
\usepackage{booktabs}
\usepackage{soul}
\usepackage{multirow}
\usepackage{subfig}
\usepackage[flushleft]{threeparttable}
\usepackage{adjustbox}
\usepackage{fancyhdr}
\usepackage[dvipsnames]{xcolor}
\usepackage[flushleft]{threeparttable}
\usepackage{tikz}
\usepackage{outlines}
\usepackage{balance}
\usepackage{tcolorbox}
\usepackage{wrapfig}
\usepackage{blindtext}
\usepackage{xcolor}
\usepackage{balance}
\usepackage{algorithm}
\usepackage{color}
\usepackage{listings}
\usepackage{cite}
\usepackage{MnSymbol,wasysym}
\usepackage{marvosym}
\usepackage[switch]{lineno}
\usepackage{xcolor,colortbl}
\usepackage{tikz}
\usepackage{graphicx}
\usepackage{stackengine}
\usepackage{color}
\usepackage{tcolorbox}
\usepackage{xcolor}
\usepackage{lineno}
\usepackage{hyperref}
\usepackage{graphicx} 
\usepackage{pifont}
\usepackage{diagbox}
\usepackage{cprotect}
\usepackage{tablefootnote}

\newcommand{\cmark}{\ding{51}}%
\newcommand{\xmark}{\ding{55}}%

\setstackEOL{\\}

\definecolor{codegreen}{rgb}{0,0.6,0}
\definecolor{codegray}{rgb}{0.5,0.5,0.5}
\definecolor{codepurple}{rgb}{0.58,0,0.82}
\definecolor{backcolour}{rgb}{0.95,0.95,0.92}

\lstdefinestyle{mystyle}{
  backgroundcolor=\color{backcolour}, commentstyle=\color{codegreen},
  keywordstyle=\color{magenta},
  numberstyle=\tiny\color{codegray},
  stringstyle=\color{codepurple},
  basicstyle=\ttfamily\footnotesize,
  breakatwhitespace=false,         
  breaklines=true,                 
  captionpos=b,                    
  keepspaces=true,                 
  numbers=left,                    
  numbersep=4pt,                  
  showspaces=false,                
  showstringspaces=false,
  showtabs=false,                  
  tabsize=1
}
\newtcolorbox{coloredalgo}{
    colback=blue!20, 
    colframe=blue!20, 
    boxrule=0pt, 
    left=-3pt, 
    right=0pt, 
    top=-2pt,
    bottom=-2pt,
}

\usepackage{tikz}
\usetikzlibrary{backgrounds}

\usepackage[dvipsnames]{xcolor}

\makeatletter
\newcommand{\algcolor}[2]{%
  \hskip-\ALG@thistlm\colorbox{#1}{\parbox{\dimexpr\linewidth-2\fboxsep}{\hskip\ALG@thistlm\relax #2}}%
}

\makeatother

\def\BibTeX{{\rm B\kern-.05em{\sc i\kern-.025em b}\kern-.08em
    T\kern-.1667em\lower.7ex\hbox{E}\kern-.125emX}}

\definecolor{Gray}{gray}{0.85}
\definecolor{LightCyan}{rgb}{0.88,1,1}

\definecolor{L1}{RGB}{248, 240, 229}
\definecolor{L2}{RGB}{234, 219, 200}

\usepackage{algorithm}
\usepackage{algpseudocode} 

\usepackage{twoopt}
\usepackage{tikz}
\usetikzlibrary{fit}

\newcommandtwoopt\Textbox[5][2.5cm][2cm]{%
\begin{tikzpicture}[remember picture,overlay]
  \coordinate (aux) at ([xshift=#1]#4);
  \node[inner ysep=5pt,yshift=0ex,draw=black,
    fit=(#3) (aux),baseline] 
    (box) {};
  \node[text width=#2,anchor=north east,
    font=\sffamily\footnotesize,align=right] 
    at (box.north east) {#5};
\end{tikzpicture}%
}
\DeclareCaptionSubType*{algorithm}

\DeclareCaptionLabelFormat{alglabel}{Alg.~#2}
    
\begin{document}
\bstctlcite{IEEEexample:BSTcontrol}

\title{GCV-Turbo: End-to-end Acceleration of \underline{G}NN-based \underline{C}omputer \underline{V}ision Tasks on FPGA 
}

\author{
\IEEEauthorblockN{ Bingyi Zhang\IEEEauthorrefmark{1}, Rajgopal Kannan\IEEEauthorrefmark{2},  Carl Busart\IEEEauthorrefmark{2}, Viktor Prasanna\IEEEauthorrefmark{1}}
\IEEEauthorblockA{
    \IEEEauthorrefmark{1}University of Southern California \IEEEauthorrefmark{2}DEVCOM Army Research Office\\
    \IEEEauthorrefmark{1}\{bingyizh, prasanna\}@usc.edu \IEEEauthorrefmark{2}\{rajgopal.kannan.civ, carl.e.busart.civ\}@army.mil}
}

\maketitle

\begin{abstract}
Graph neural networks (GNNs) have recently empowered various novel computer vision (CV) tasks. In GNN-based CV tasks, a combination of CNN layers and GNN layers or only GNN layers are employed. This paper introduces GCV-Turbo, a domain-specific accelerator on FPGA for end-to-end acceleration of GNN-based CV tasks. GCV-Turbo consists of two key components: (1) a \emph{novel} hardware architecture optimized for the computation kernels in both CNNs and GNNs using the same set of computation resources. (2) a PyTorch-compatible compiler that takes a user-defined model as input, performs end-to-end optimization for the computation graph of a given GNN-based CV task, and produces optimized code for hardware execution. The hardware architecture and the compiler work synergistically to support a variety of GNN-based CV tasks. We implement GCV-Turbo on a state-of-the-art FPGA and evaluate its performance across six representative GNN-based CV tasks with diverse input data modalities (e.g., image, human skeleton, point cloud). Compared with state-of-the-art CPU (GPU) implementations, GCV-Turbo achieves an average latency reduction 
 of $68.4\times$ ($4.1\times$) on these six GNN-based CV tasks. Moreover, GCV-Turbo supports the execution of the standalone CNNs or GNNs, achieving performance comparable to that of state-of-the-art CNN (GNN) accelerators for widely used CNN-only (GNN-only) models.
\end{abstract}


\section{Introduction}
\label{sec:introduction}

\begin{figure}[ht]
     \centering
     \includegraphics[width=8cm]{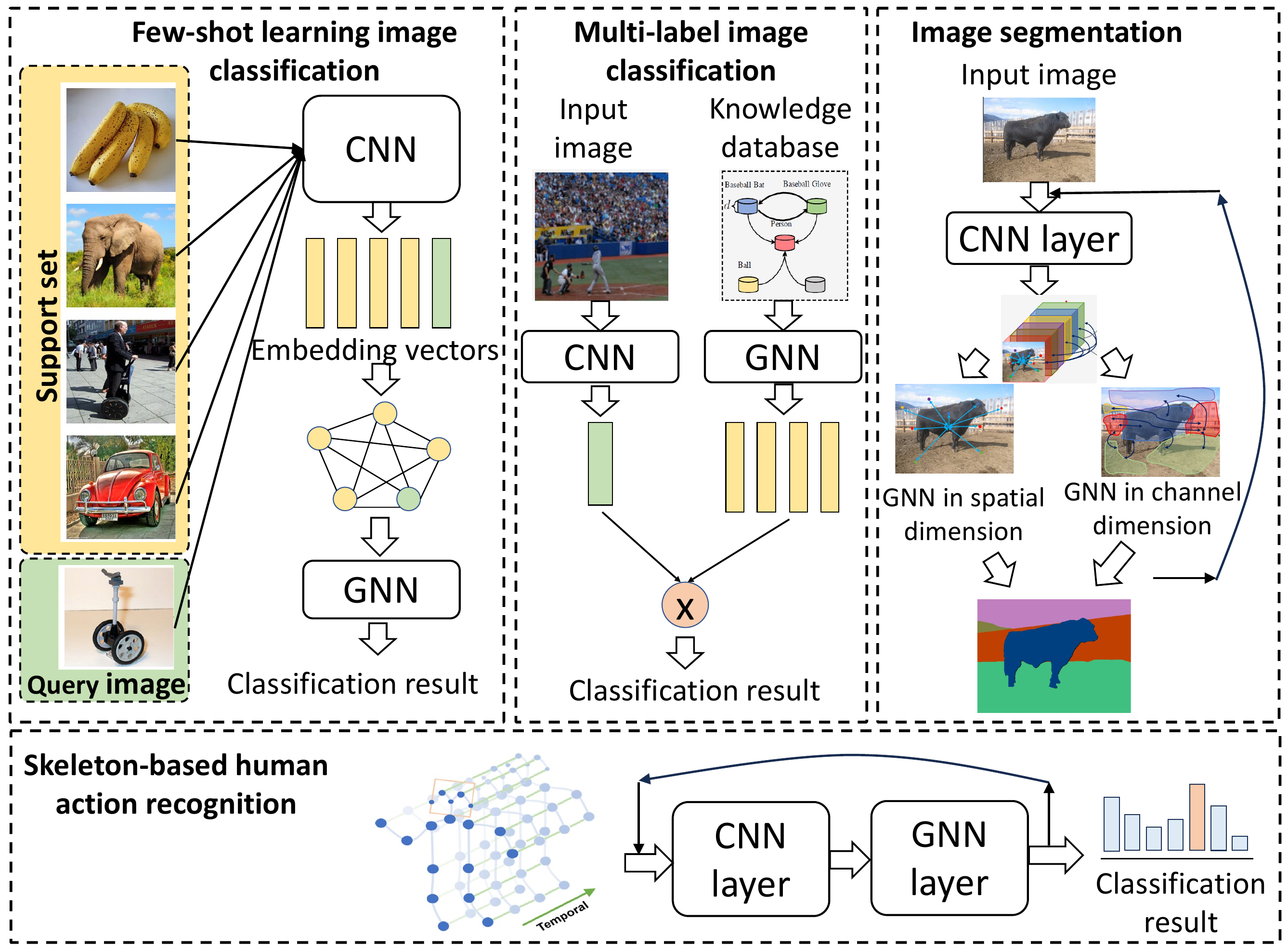}
     \vspace{-0.2cm}
     \caption{Examples of GNN-based CV tasks \cite{garcia2018few, chen2019multi, zhang2019dual, yan2018spatial}}
     \vspace{-0.2cm}
     \label{fig:examples}
\end{figure}

Graph Neural Networks (GNNs)  are playing an increasingly important role in various computer vision (CV) tasks \cite{chen2022survey, chowdhury2023state}. Figure \ref{fig:examples} demonstrates several examples. These applications utilize the combined power of convolution in CNN layers and message passing in GNN layers. 
This has given rise to a new domain called GNN-based CV:  
\emph{CV tasks that utilize  a combination of CNN and GNN layers (e.g., iteratively interleaving CNN layer and GNN layer) or rely solely on GNN layers.} 

GNN layers have gained widespread adoption in CV tasks because:
(1) Firstly, GNN layers facilitate \emph{label-efficient} image classification. Training standalone CNN \cite{he2016deep} or vision transformer (ViT) \cite{dosovitskiyimage}  
typically requires a substantial number of labeled images. For instance, achieving high accuracy with ViTs requires over 300 million labeled images. In contrast, researchers have devised label-efficient few-shot learning techniques \cite{garcia2018few} that combine GNN layers and CNN layers, requiring only a small number of labeled images.
(2) Secondly, GNN layers can naturally handle \emph{non-Euclidean} data structures in diverse CV tasks, such as point clouds \cite{shi2020point, qi2017pointnet, qi2017pointnet++}, 3D meshes \cite{wen2021dual, tang2023feature}. In contrast, the convolution of the CNN layer and the multi-head self-attention (MSA) of ViTs are designed for regular grids and cannot be directly employed with non-Euclidean data structures. For example, convolution operates on 2D grids, and MSA \cite{dosovitskiyimage} relies on positional encodings on 2D grids.
(3)  Thirdly, the message passing of GNN layers excel in \emph{relation learning} for various CV tasks, allowing them to understand complex object relationships. In video action recognition, a CNN \cite{redmon2016you} detects multiple objects, while GNN layers \cite{wu2019learning} are employed to capture object relationships.

Given this domain's expanding scope and future relevance, there is an urgent need for end-to-end acceleration for these GNN-based CV tasks. For example, in autonomous driving, low latency inference is crucial to ensure safety.  Nevertheless, it poses significant challenges:  while there are various CNN accelerators \cite{xilinxdpu, 9065523, jouppi2018motivation, abdelfattah2018dla, yu2019opu, chen2014dadiannao, 7480791} or GNN accelerators \cite{yan2020hygcn, geng2020awb, zhang2021boostgcn, sarkar2023flowgnn, liang2020deepburning, zhang2020hardware, geng2021gcn, li2021gcnax} proposed. The hardware architecture of these CNN or GNN accelerators is optimized solely for one layer type, which is inefficient for the end-to-end acceleration of GNN-based CV. For example, CNN accelerators \cite{xilinxdpu, 9065523, jouppi2018motivation, abdelfattah2018dla, yu2019opu, chen2014dadiannao, 7480791} are not efficient for message passing in GNN layers while GNN accelerators achieve suboptimal performance on convolution operation of CNN layers. While we can potentially combine a CNN accelerator and a GNN accelerator for GNN-based CV, it can lead to sub-optimal performance due to resource underutilization. For example, when executing a CNN layer, the GNN accelerator will be idle and vice versa. Another possible solution is to build FPGA bitstreams for CNN and GNN layers, respectively. However, this requires  dynamic reconfiguration of FPGA for executing a model, which can incur significant latency and is not suitable for latency-sensitive applications. \textbf{(2)} GNN-based CV model has a mixture of dataflow because the GNN layer and CNN layer can be interleaved but have very different data layouts. Existing compilers and hardware architectures of the aforementioned CNN or GNN accelerators are not optimized for this dataflow mixture, which can potentially lead to large overhead in transforming data layouts between two types of layers. Coordinating the data layout between the CNN and GNN layers requires non-trivial compiler-hardware codesign. \textbf{(3)} General purpose processors (CPU, GPGPU) are not well-suited for low-latency inference of GNN-based CV. Because they have complex cache hierarchies leading to large and unpredictable memory access latency, unsuitable for latency-sensitive applications.
Given that there are no existing accelerators for GNN-based CV, the execution of existing GNN-based CV \cite{garcia2018few, chen2019multi, zhang2019dual, yan2018spatial, zhang2023graph, qi2017pointnet}  rely on CPU/GPU, leading to suboptimal performance. \textbf{(4)} Moreover, autonomous driving systems execute various CV tasks including non-GNN CV. Therefore, an accelerator should not only achieve high performance for GNN-based CV, but also not sacrifice much performance for tasks that utilize standalone CNNs or GNNs.

To address the above challenges, we propose GCV-Turbo, a domain-specific accelerator on FPGA for end-to-end acceleration of GNN-based CV. 
Unlike existing CNN and GNN accelerators, the architecture design of GCV-Turbo employs the \emph{resource sharing strategy} that different computation kernels in CNNs and GNNs share the same computation resources for improved resource utilization. 
Moreover, the compiler not only optimizes CNN layers or GNN layers but also performs end-to-end optimizations for the mixture dataflow of CNN and GNN layers. 
Our main contributions are:
\begin{itemize}
    \item We propose GCV-Turbo, the \emph{first} domain-specific accelerator for end-to-end acceleration of GNN-based CV tasks.
    \item We design a novel hardware architecture with a flexible data path and memory organization capable of executing various computation kernels in CNN and GNN layers using the \emph{same} set of hardware resources.
    \item We develop a customized compiler for end-to-end optimizations that reduces inference latency of GNN-based CV, including (1) optimizations for data manipulation between CNN layers and GNN layers, (2) data layout centric mapping, (3) sparsity-aware computation primitive mapping. 
    \item  We implement the hardware design on a state-of-the-art FPGA board, Alveo U250.  Evaluated on six representative GNN-based CV tasks, GCV-Turbo achieves average latency reduction of $68.4\times$ and $4.1\times$  compared with the state-of-the-art implementations on CPU and GPU, respectively.
    \item We compare GCV-Turbo with state-of-the-art CNN and GNN accelerators. GCV-Turbo demonstrates performance comparable to CNN DSAs for CNN-only models (with a speedup of $0.88$ to $0.93 \times$), and to GNN accelerators for GNN-only models (with a speedup of $1.03$ to $1.25 \times$).
\end{itemize}
To the best of our knowledge, GCV-Turbo is the first hardware-compiler codesign capable of executing both CNN and GNN layers, optimized for the end-to-end acceleration of GNN-based CV, and also maintaining good performance on tasks that utilize standalone CNN or GNN.

\section{Background}
\label{sec:background}

\subsubsection{GNN-based Computer Vision Tasks}
\label{subsec:GNN-based-CV}

\begin{figure}[ht]
     \centering
     \vspace{-0.1cm}
     \includegraphics[width=8cm]{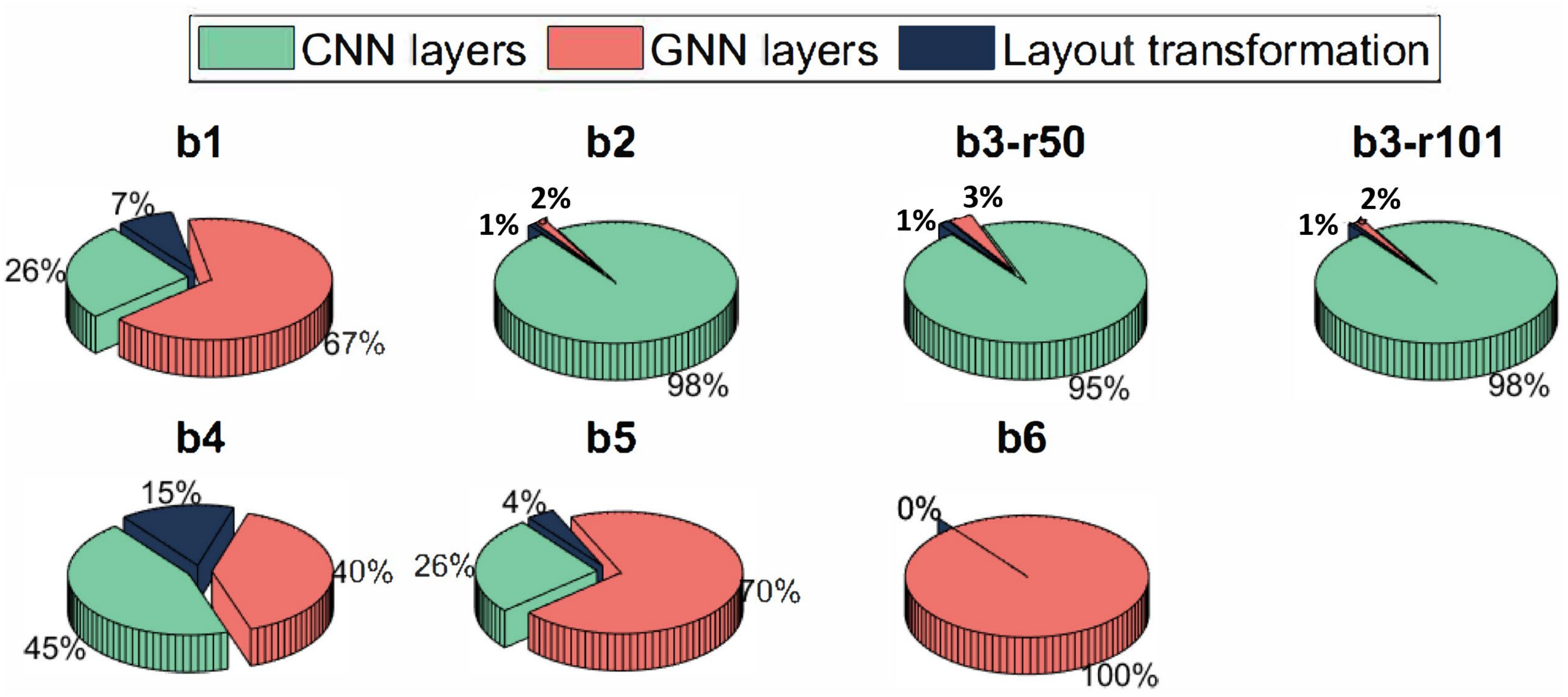}
     \vspace{-0.1cm}
     \cprotect\caption{Breakdown analysis of GNN-based CV tasks (\verb|b1|-\verb|b6|) on state-of-the-art GPU (RTX A5000). 
 The details of the models and datasets are elaborated in Section \ref{sec:implementation}.}
     \vspace{-0.4cm}
     \label{fig:pie-gpu}
\end{figure}

Figure \ref{fig:examples} shows several representative GNN-CV tasks. We conduct an experimental study to understand the challenges of accelerating GNN-based CV: (1) In CNN-based CV tasks, both CNN and GNN layers can be computationally extensive. Moreover, the computation workloads of CNN/GNN layers vary in tasks ranging from $2\%-100\%$ (Figure \ref{fig:pie-gpu}). Directly combining a CNN accelerator and a GNN accelerator can lead to severe hardware underutilization. For example, the GNN accelerator will be idle when executing a CNN layer. This underutilization can increase the inference latency. (2) In GNN-based CV tasks, the CNN layer and GNN layer have very different data layouts for input and output data. Moreover, the CNN layer and GNN layer can be interleaved (which is for better feature fusion in GNN-based CV. See image segmentation and skeleton-based human action recognition in Figure \ref{fig:examples}). Switching the data layout (including permute(), transpose(), and other indexing functions) between the CNN layer and the GNN layer can lead to significant overhead, taking $1\%-15\%$ execution time on a state-of-the-art GPU platform (Figure \ref{fig:pie-gpu}). This can be more severe on embedded platforms with limited memory bandwidth since layout transformation is memory-bound.  (3) General purpose processors (CPU, GPU) are hard to achieve low-latency inference for GNNs \cite{yan2020hygcn, sarkar2023flowgnn, zhang2021boostgcn}, because GNN has irregular data access patterns and memory access patterns. Due to the complex cache hierarchy, CPU and GPU have low efficiency \cite{yan2020hygcn, sarkar2023flowgnn, zhang2021boostgcn} for executing GNNs.

\subsubsection{Domain Specific Accelerators}
\label{subsec:Domain-specific-Accelerators}
A multitude of domain-specific accelerators (DSAs) \cite{xilinxdpu, 9065523, jouppi2018motivation, abdelfattah2018dla, yu2019opu, chen2014dadiannao} have been proposed to accelerate CNNs. However, these CNN-specific DSAs have challenges in both hardware design and compiler implementation when it comes to executing GNN-based CV tasks, as elaborated in Section \ref{sec:introduction}. Recently, a DSA known as GraphAGILE \cite{zhang2023graphagile} has emerged to accelerate GNNs. Unfortunately, the GraphAGILE compiler does not accommodate CNNs, which is crucial for GNN-based CV tasks. Moreover, GraphAGILE does not explore the data sparsity in GNNs, which can result in suboptimal performance when applied to GNN-based CV tasks.
Meanwhile, there exist other accelerators \cite{wen2021rfc} designed for specific GNN-based CV tasks. RFC-HyPGCN \cite{wen2021rfc} specializes in running 2S-AGCN \cite{shi2019two} for human-skeleton-based action recognition, while Pointacc \cite{lin2021pointacc} is tailored to accelerate several GNNs utilized in point cloud applications. In summary, prior research efforts either (1) design DSAs exclusively for CNNs or GNNs, or (2) design accelerators optimized for specific GNN-based CV tasks.



\textbf{}
\section{Overview}
\label{sec:overview}

\subsection{Problem Definition}
\label{subsec:Problem-Definition}
Our objective is to perform end-to-end  \emph{inference} acceleration of GNN-based CV tasks. End-to-end acceleration refers to reducing the  inference latency of a GNN-based CV task, which is duration from when the input data is given to the time when the inference result is obtained. 
This includes data loading from external memory, executing all the layers of the model on the accelerator, and storing the results in the external memory.
To this end, we propose a compiler-hardware codesign. The compilation is an \emph{offline} process. The GCV-Turbo compiler takes a user-defined model (written in PyTorch \cite{lerer2019pytorch} and PyTorch Geometric \cite{fey2019fast}) as input and generates optimized code for hardware execution.
The GCV-Turbo hardware design has a fixed architecture that execute various models without reconfiguring the FPGA. This is important for many real-world systems, such as autonomous driving, which execute various models for various data modality. We target \emph{latency-sensitive} applications such as autonomous driving, where inference latency should be low to ensure safety.

\subsection{Overview of GCV-Turbo}
\label{subsec:overview-GCV-Turbo}
\begin{figure}[ht]
     \centering
     \includegraphics[width=6.5cm]{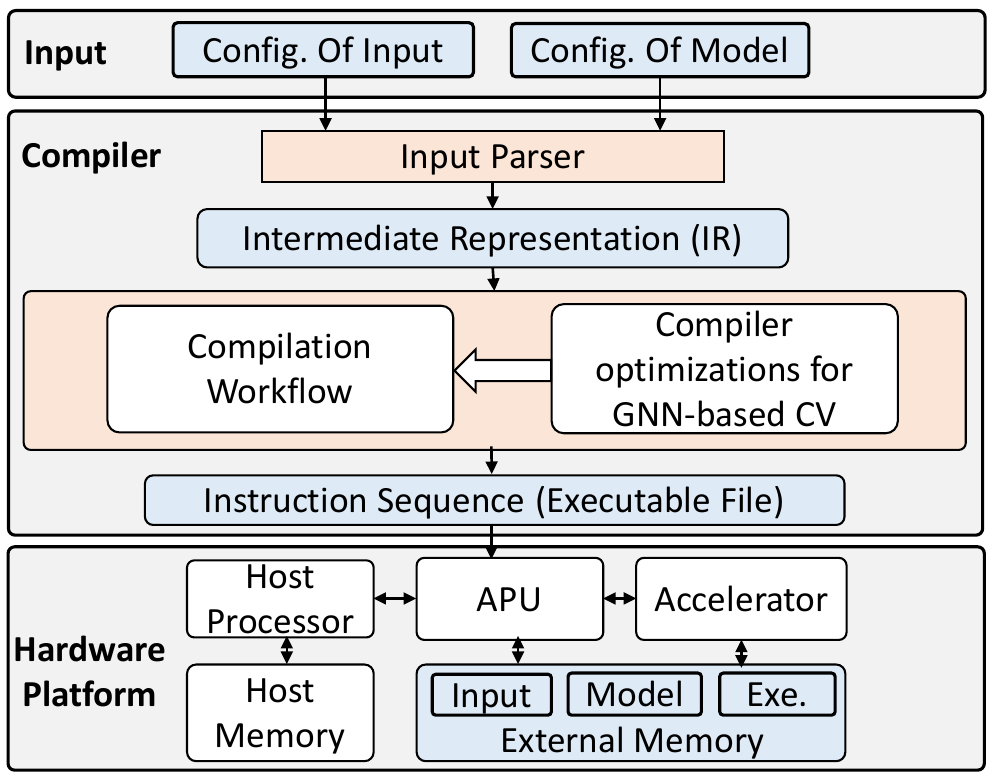}
     \vspace{-0.2cm}
     \caption{Overview of GCV-Turbo}
     \label{fig:GCV-Turbo-Overview}
     \vspace{-0.5cm}
\end{figure}

Figure \ref{fig:GCV-Turbo-Overview} illustrates the overview of GCV-Turbo:
(1) \emph{Compiler}: It is executed on the host processor. The \emph{input parser} generates the intermediate representation  (computation graph) from the given  input model. The computation graph or intermediate representation is the high-level representation of the input model with each node representing a layer and each arrow representing the data dependency. Then, compiler performs five-step compilation to map the input model onto the hardware accelerator. We apply several compiler optimizations (Section \ref{subsec:Compiler-Optimization}) for GNN-based CV. Finally, the compiler generates an instruction sequence for hardware execution.
(2) \emph{Application processing unit (APU)}: The APU of FPGA \cite{Microblaze-link} takes the instruction sequence  as input and launches the workload of inference on the hardware accelerator. 
(3) \emph{Hardware accelerator}: The hardware accelerator executes the computation tasks scheduled by APU. 


 \emph{Hardware design:} As discussed in Section \ref{sec:introduction}, existing CNN or GNN accelerators suffer from inefficiency when handling GNN-based CV tasks. To tackle this challenge, we identify fundamental computation primitives (Section \ref{subsec:comp-primitives}) capable of representing computation kernels in both GNNs and CNNs. Subsequently, we design a flexible data path and memory organization for efficient execution of these computation primitives within our hardware design. This enables our accelerator to support both CNNs and GNNs.
Meanwhile, our proposed accelerator incorporates an instruction set (Section \ref{subsec:instruction-set}) providing software-like programmability. Note that our hardware design employs \emph{resource sharing strategy} (Section \ref{sec:Hardware-arch}) such that the computation kernels of CNN and GNN share the same set of computation resources. 

 \begin{figure*}[ht]
     \centering
     \includegraphics[width=18cm]{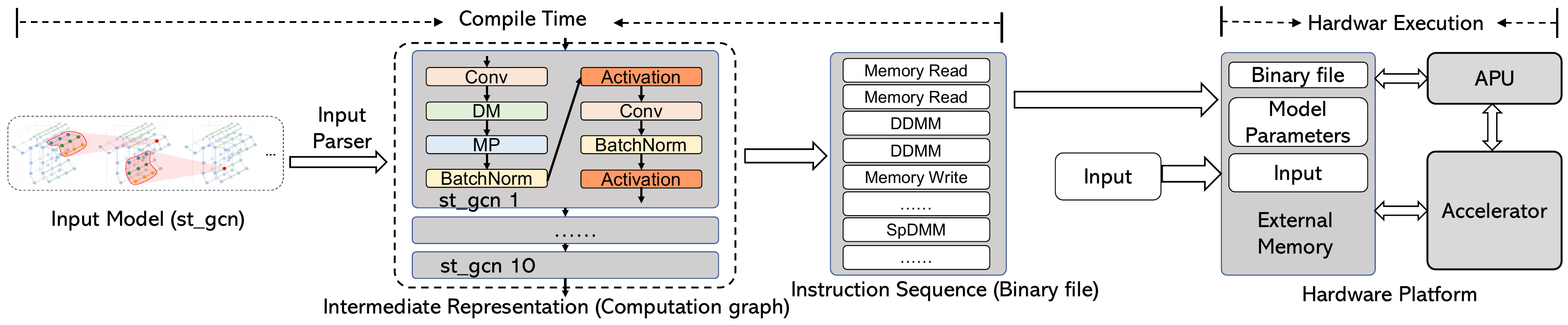}
     \vspace{-0.1cm}
     \caption{Workflow of GCV-Turbo using the skeleton-based human action recognition \cite{yan2018spatial} as an example. }
     \label{fig:GCV-Turbo-workflow}
     \vspace{-0.2cm}
\end{figure*}

\emph{Compiler design:} Designing a compiler to support GNN-based CV is not merely merging separate compiler optimization for CNNs and GNNs. Instead, it needs the end-to-end optimization of the computation graph of a GNN-based CV model. Because:
(1) a GNN-based CV task often comprises both CNN and GNN layers, and these layers can be interwoven (e.g., \cite{yan2018spatial}).
(2) These two layer types exhibit different data layouts and memory access patterns. Without careful dataflow optimization, switching data layouts can lead to substantial overhead and increased memory access latency.
To address this challenge, we devise a five-step compilation workflow (Section \ref{sec:compiler}) with various compiler optimizations for GNN-based CV (Section \ref{subsec:Compiler-Optimization}).

\emph{Workflow}: The workflow is illustrated in Figure \ref{fig:GCV-Turbo-workflow}. At \emph{compile time}, the compiler takes the user-defined model as input and produces the intermediate representation (i.e., computation graph). 
The compiler then performs a five-step compilation to generate an instruction sequence stored in a binary file. During hardware execution, the APU reads the binary file and schedules the computation tasks on the hardware accelerator.

\emph{Experimental study}: We conduct the comprehensive experimental study on six representative GNN-based CV tasks (Section \ref{subsec:benchmark-baseline}). Because these tasks (1) cover various use cases and data modalities (See Table \ref{tab:benchmark-details}) in real-world applications, such as autonomous driving, (2) cover various computational characteristics of GNN-based CV (See Figure \ref{fig:pie-gpu}), such as varying portions of CNN/GNN layers, varying patterns of layout transformation between CNN and GNN layers. Evaluating these tasks, we expect GCV-Turbo to perform similarly on a broad range of GNN-based CV tasks.



 \begin{figure*}[ht]
     \centering
     \includegraphics[width=16.5cm]{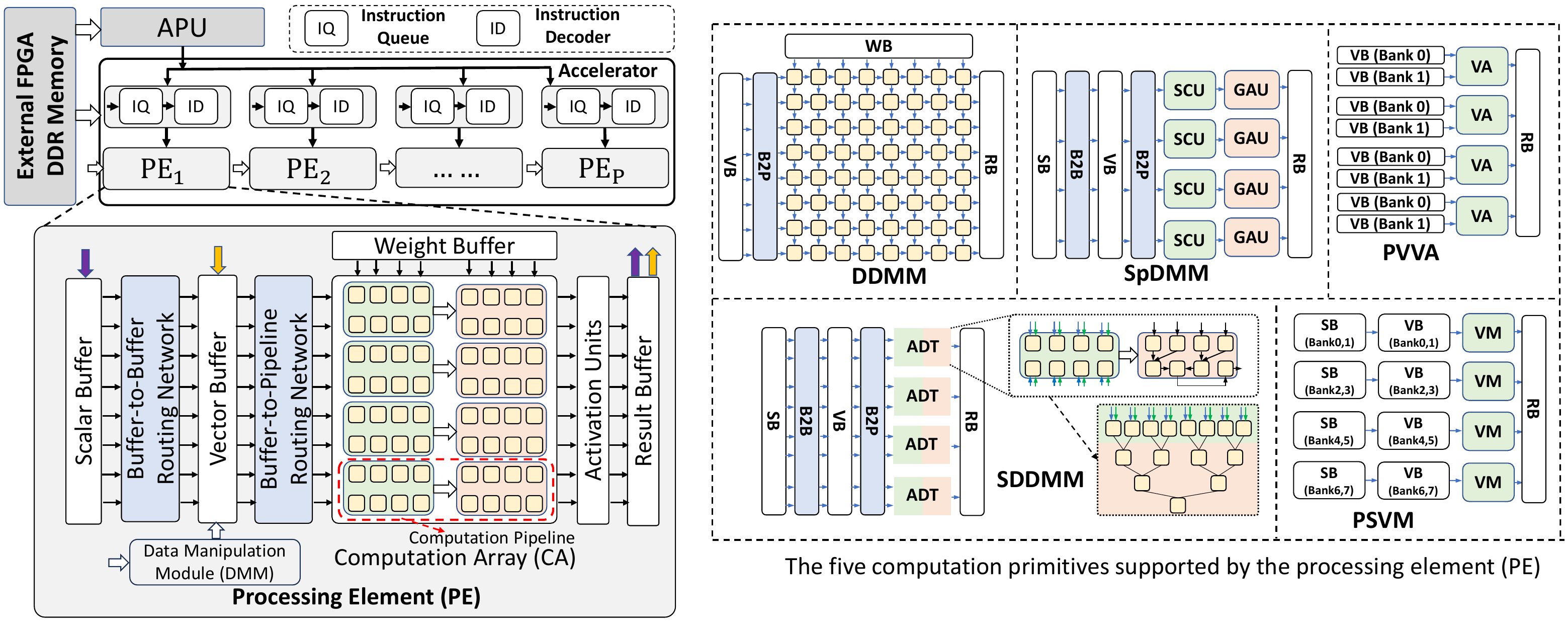}
     \vspace{-0.2cm}
     \caption{Architecture  of hardware accelerator, and the basic computation primitives supported by a PE.}
     \vspace{-0.3cm}
     \label{fig:hard-arch}
\end{figure*}

\section{Hardware Architecture}
\label{sec:Hardware-arch}

As illustrated in Figure \ref{fig:hard-arch}, GCV-Turbo has a unified hardware architecture that efficiently executes various computation primitives (Section \ref{subsec:comp-primitives}) in CNNs and GNNs. The accelerator has multiple parallel processing elements (PEs), with each having an Instruction Queue (IQ) and an Instruction Decoder (ID). Each PE has a computation array (CA) with $p_{\text{ca}}^{2}$ computation units. Each computation unit executes the basic arithmetic operations. Each PE has a Scalar Buffer (SB), a Vector Buffer (VB), a Weight Buffer (WB), and a Result Buffer (RB). Each Buffer (SB/VB/WB/RB) has $p_{\text{ca}}$ memory banks. Each bank can output $p_{\text{ca}}$ data per cycle. 
There are two all-to-all data routing networks  -- Buffer-to-Buffer (B2B) and Buffer-to-Pipeline (B2P) Routing Network. 
Data Manipulation Module performs transformations of data layout between different layers (e.g., CNN layer and GNN layer). 

\emph{Resource sharing}: Note that in a PE, different computation primitives share the same set of computation units, data buffers, and routing networks. See more details in Section \ref{subsec:comp-primitives}. This increases the resource utilization for executing a GNN-based CV task. For resource sharing, it only requires extra extra wire connections and hardware multiplexers for selecting data path for different computation primitives, which incur small hardware cost (See Section \ref{sec:implementation}).

\subsection{Computation Primitives}
\label{subsec:comp-primitives}

In GNN-based CV tasks,  we identify five basic computation primitives (Figure \ref{fig:hard-arch}), including \emph{dense-dense matrix multiplication (DDMM)}, \emph{sparse-dense matrix multiplication (SpDMM)}, \emph{sampled dense-dense matrix multiplication (SDDMM)}, \emph{parallel scalar-vector multiplication (PSVM)}, and \emph{parallel vector-vector addition (PVVA)}. 
Each layer can be mapped to these basic computation primitives.
The PE has a flexible architecture to support these computation primitives. 
Each PE maintains  hardware multiplexers to select the data path for executing various primitives. Switching among primitives incurs one clock cycle overhead. For simplicity, the input to a computation primitive are two matrices denoted as $\mathbf{X} \in \mathbb{R}^{s_{1} \times s_{2}}$ and $ \mathbf{Y} \in \mathbb{R}^{s_{2} \times s_{3}}$. The output matrix is denoted as $\mathbf{Z} \in \mathbb{R}^{s_{1} \times s_{3}}$.

\vspace{0.05cm} 
\noindent \textbf{DDMM}: DDMM executes $\mathbf{X} \times \mathbf{Y}$, and views $\mathbf{X}$ and $\mathbf{Y}$ as dense matrices. To this end, the computation array is organized as a 2-D systolic array (See Figure \ref{fig:hard-arch}) with localized interconnection. $\mathbf{X}$ and $\mathbf{Y}$ are stored in VB and WB, respectively. Different from traditional 2-D systolic arrays, DDMM incorporates a B2P routing network for shuffling the position of input vectors (rows of $\mathbf{X}$), which supports data layout transformation between CNN layer and GNN layer (See Section \ref{subsec:Compiler-Optimization}). DDMM can execute $p_{\text{ca}} \times p_{\text{ca}}$ multiply-accumulate (MAC) operations in each clock cycle. 



\noindent \textbf{SpDMM}: SpDMM executes $\mathbf{X} \times \mathbf{Y}$ where $\mathbf{X}$ is a sparse matrix. The computation array is organized as multiple pipelines with each having a Scatter Unit (SCU) and a Gather Unit (GAU). Each non-zero element in $\mathbf{X}$ is represented using a three-tuple $(src, dst, val)$, denoting row index, column index, and  value, respectively. The execution follows the scatter-gather paradigm \cite{zhou2019hitgraph, chen2021thundergp} as shown in Algorithm \ref{alg:SpDMM-scatter-gather}. Executing $\mathbf{X} \times \mathbf{Y}$ takes $l_{\text{SpDMM}}$ clock cycles:
$
    \label{eq:SpDMM} 
    {\textstyle l_{\text{SpDMM}}(\mathbf{X}, \mathbf{Y}) = \lceil\frac{Nonz(\mathbf{X})}{p_{\text{ca}}/2} \rceil \times \lceil \frac{s_{3}}{p_{\text{ca}}} \rceil}
$
where $\textstyle{ Nonz(\mathbf{X})}$ denotes the number of non-zeros in $\mathbf{X}$.

\begin{algorithm}
\caption{SpDMM using Scatter-Gather paradigm}\label{alg:SpDMM-scatter-gather}
\begin{small}
\begin{algorithmic}

\While {not done}  {\color{blue}\Comment{Pipelined Execution}}
\For{each  $(src,dst,val)\in \mathbf{X}$ in SB}  {\color{blue}\Comment{Data Fetching}}
\State Route $(src,dst,val)$ from SB to VB {\color{blue}\Comment{B2B}}
\State Fetch  row $src$ of $\mathbf{Y}$: $\mathbf{Y}[src]$ from VB
\State Form input pair \{$\mathbf{Y}[src]$, $(src,dst,val)$\} 
\State Route the input pair to pipeline $dst\%(p_{\text{ca}}/2)$
{\color{blue}\Comment{B2P}}
\EndFor 
\For{each input pair \{$\mathbf{Y}[src]$, $(src,dst,val)$\}  } 
\State Produce $\mathbf{u} \gets $ $val \times \mathbf{Y}[src]$ {\color{blue}\Comment{Scatter Unit (SCU)}}
\State {Update $\mathbf{Z}[dst] += \mathbf{u}$ } {\color{blue}\Comment{Gather Unit (GAU)}}
\EndFor
\EndWhile
\end{algorithmic}
\end{small}
\end{algorithm}

\noindent \textbf{SDDMM}: SDDMM executes $\mathbf{Z} = \mathbf{A} \odot (\mathbf{X}\mathbf{Y})$ $(  \mathbf{A} \in \mathbb{R}^{s_{1} \times s_{3}})$, where $\odot$ is the element-wise multiplication. $\mathbf{A}$ is a sampling matrix where each element is either $1$ or $0$ to sample results from $\mathbf{X}\mathbf{Y}$. For example, if $\mathbf{A}[i][j] = 1$, then $\mathbf{Z}[i][j] = \langle \mathbf{X}[i], \mathbf{Y}[j]\rangle$ where $\langle,\rangle$ denotes vector inner product operator. If $\mathbf{A}[i][j] = 0$, $\mathbf{Z}[i][j]=0$. Each computation pipeline is organized as an adder tree (ADT). The execution of SDDMM is shown in Algorithm \ref{alg:SDDMM}. Executing $\mathbf{A} \odot (\mathbf{X}\mathbf{Y})$ takes $l_{\text{SDDMM}}$ clock cycles, where
$
    {\textstyle l_{\text{SDDMM}}(\mathbf{X}, \mathbf{Y}) = \lceil\frac{Nonz(\mathbf{X})}{p_{\text{ca}}/2} \rceil \times \lceil \frac{s_{2}}{p_{\text{ca}}} \rceil}
$.

\begin{algorithm}
\caption{Sampled dense-dense matrix multiplication}\label{alg:SDDMM}
\begin{small}
\begin{algorithmic}
\While {not done} {\color{blue}\Comment{Pipelined Execution}}
\For{each  $(src,dst)\in \mathbf{A}$ in SB}  {\color{blue}\Comment{Data Fetching}}
\State Route $(src,dst)$ from SB to VB {\color{blue}\Comment{B2B}}
\State Fetch $\mathbf{X}[src]$ and $\mathbf{Y}[dst]$ from VB
\State Form input pair \{$\mathbf{X}[src]$, $\mathbf{Y}[dst]$\} 
\State Route the input pair to a pipeline
{\color{blue}\Comment{B2P}}
\EndFor 
\For{each input pair} {\color{blue}\Comment{Computation}}
\State {Update $\mathbf{Z}[src][dst] += \langle \mathbf{X}[src], \mathbf{Y}[dst]\rangle$ } {\color{blue}\Comment{ADT}}
\EndFor
\EndWhile
\end{algorithmic}
\end{small}
\end{algorithm}

\noindent \textbf{PSVM}: To execute PSVM, the computation array is organized as $p_{\text{ca}}/2$ independent pipelines. Each pipeline has a vector multiplier (VM) to execute the multiplication between a scalar and a vector of length $p_{\text{ca}}$. A PE can execute $p_{\text{ca}}^{2}/2$ multiply operations per clock cycle. PSVM can be used to perform matrix-vector multiplication.

\noindent \textbf{PVVA}: To execute PVVA, for $p_{\text{ca}}/2$ independent pipelines, each pipeline has a  vector adder (VA) to execute the vector addition between two vectors of length $p_{\text{ca}}$. A PE can execute $p_{\text{ca}}^{2}/2$ addition operations per cycle. PVVA can be used to execute matrix addition.

\subsection{Instruction Set}
\label{subsec:instruction-set}
We develop a customized instruction set, including computation instructions, memory read/write instructions. 
(1) \emph{Computation Instructions} includes the instruction for each computation primitives (e.g., DDMM instruction). Each instruction contains the meta data (e.g., matrix size) of the corresponding computation primitive. The Instruction Decoder decodes the instruction and generates control signal for the PE to execute the computation primitives in pipelined manner.
\emph{Memory Read/Write Instructions} launch the data transactions between the on-chip buffer  and the external memory.

\section{Compiler}
\label{sec:compiler}

Existing compilers for CNN or GNN accelerator \cite{xilinxdpu, 9065523, jouppi2018motivation, abdelfattah2018dla, yu2019opu, chen2014dadiannao, 7480791, ref-graphagile} support only one type of model (CNN or GNN). In contrast, GCV-Turbo offers an end-to-end compilation/optimization workflow for GNN-based CVs.
For a given input model developed using PyTorch, the Input Parser converts it into an intermediate representation (Section \ref{subsec:IR-and-CP}), which serves as the computation graph underlying the inference process.
The compiler then performs a five-step compilation (Section \ref{subsec:Compilation-Workflow}) to generate an instruction sequence. Especially, we perform a number of specific optimizations  (Section \ref{subsec:Compiler-Optimization}) for GNN-based CV tasks: including (1) data manipulation (DM) layer generation, (2) layer fusion for DM layer, (3) uniform mapping, (4) data layout centric mapping, and (5) sparsity-aware primitive mapping. Our compiler utilizes the infrastructure of TVM framework \cite{chen2018tvm}. Based on it, we develop our own input parser, intermediate representation, compilation workflow, and compiler optimizations.

\subsection{Intermediate Representation}
\label{subsec:IR-and-CP}

We develop the intermediate representation (IR) for the following set of computation layers in GNN-based CV tasks:

\emph{Convolutional (Conv) Layer}: The input $\mathfrak{F}_{\text{in}}$ has $c_{\text{in}}$ feature maps (channels), each having a size of $h_{\text{in}}\times w_{\text{in}}$. 
The output $\mathfrak{F}_{\text{out}}$  has  $c_{\text{out}}$ feature maps  (channels) with each having the size of $h_{\text{out}} \times w_{\text{out}}$. The convolution kernel $\mathfrak{W}$ has the size of $c_{\text{out}} \times c_{\text{in}} \times k_{1} \times k_{2}$. The output $\mathfrak{F}_{\text{out}}$ is obtained through 2D convolution between input $\mathfrak{F}_{\text{in}}$  and kernel $\mathfrak{W}$.

\emph{Message Passing (MP) Layer}: It is used in GNNs for message passing within graph $\mathcal{G}(\mathcal{V}, \mathcal{E})$.  The input are vertex feature vectors $\{\mathbf{h}_{\text{in}}[v]\in \mathbb{R}^{f}:v\in \mathcal{V}\}$ and edges $\{e_{vu}\in \mathbb{R}^{1}:e_{vu} \in \mathcal{E}\}$. The output vertex feature vectors $\{\mathbf{h}_{\text{out}}[v]\in \mathbb{R}^{f}:v\in \mathcal{V}\}$ are obtained through message passing:
$
    {\textstyle \mathbf{h}_{\text{out}}[v] = \rho(\{e_{uv} \cdot \mathbf{h}_{\text{in}}[u]: u\in \mathcal{N}(v)\})}
$
where $\mathcal{N}(v)$ denotes the set of neighbors of $v$, and  $\rho()$ is the element-wise reduction function, such as $\text{Max}()$ and $\text{Sum}()$.

\emph{Linear Layer}: In a Linear Layer, an input matrix $\mathbf{H}^{\text{in}}$ is multiplied by a weight matrix $\mathbf{W}$ to obtain output matrix $\mathbf{H}^{\text{out}}$.

\emph{Vector Inner Product (VIP) Layer}:  The inputs are the vertex feature vectors $\{\mathbf{h}_{\text{in}}[v]\in \mathbb{R}^{f}:v\in \mathcal{V}\}$, and predefined edge connectivity $\{e_{vu}\in \mathbb{R}^{1}:e_{vu} \in \mathcal{E}\}$ with the value of $e_{vu}$ to be calculated. $e_{vu}$ is calculated by:
$
    e_{uv} = \langle \mathbf{h}_{\text{in}}[u], \mathbf{h}_{\text{in}}[v]\rangle
$
where $\langle, \rangle$ denotes vector inner product.

Data Manipulation (DM) Layers: The DM layer is our proposed new layer that represents the necessary data manipulation operation between the CNN layer and the GNN layer. See details in Section \ref{subsubsec:dm-gen}.

\emph{Other Layers}: Include other types of layers, such as Pooling layers, Normalization (Norm) layers, and Activation layers.

Following the convention of TVM \cite{chen2018tvm}, we implement the IR of each layer as a tensor IR function (\verb|T.prim_func|) using TVMScript. The input parser of the compiler generates the computation graph from the input model and represents each layer using the IR.


\subsection{Compilation Workflow}
\label{subsec:Compilation-Workflow}
We introduce the basic compilation workflow of GCV-Turbo, which has five steps:
\begin{itemize}
    \item \emph{Step 1 - layer fusion}: \label{subsubsec:step-1} For the computation graph of an input model, the layer fusion step merges some layers (e.g., activation layer, normalization layer) into the adjacent layers to facilitate task-level parallelism, reduce memory traffic, and reduce overall computation complexity. 
    \item \emph{Step 2 - layer-to-matrix operation mapping}: \label{subsubsec:step-2} For each layer in the computation graph, the compiler maps it into a set of matrix operations (e.g., matrix multiplication). 
    \item \emph{Step 3 - data tiling and task partitioning}:  \label{subsubsec:step-3} Because the accelerator has limited on-chip memory, this step performs data tiling for each matrix operation. Therefore, a large matrix operation can be decomposed into a set of matrix operations on small data tiles.
    \item \emph{Step 4 - mapping matrix operation to Computation Primitive}: This step maps each matrix operation into the basic computation primitives (Section \ref{subsec:comp-primitives}) that are supported by the accelerator. 
    \item \emph{Step 5 - Task scheduling}: This step plans the execution of the computation graph on the accelerator. The proposed accelerator processes the model layer-by-layer. For each layer, the APU schedules its computation using a centralized load balancing scheme \cite{barney2010introduction} for workload balance between PEs, according the status (idle or busy) of PEs.
\end{itemize}
In our design, each step is implemented as a compilation pass. Finally, the compiler generates an instruction sequence for hardware execution.

\subsection{Compiler Optimizations for GNN-based CV tasks}
\label{subsec:Compiler-Optimization}
We introduce the following set of compiler optimizations for GNN-based CV:

\subsubsection{Data Manipulation Layer Generation} \label{subsubsec:dm-gen} 
\begin{figure}[ht]
     \centering
     \includegraphics[width=6.5cm]{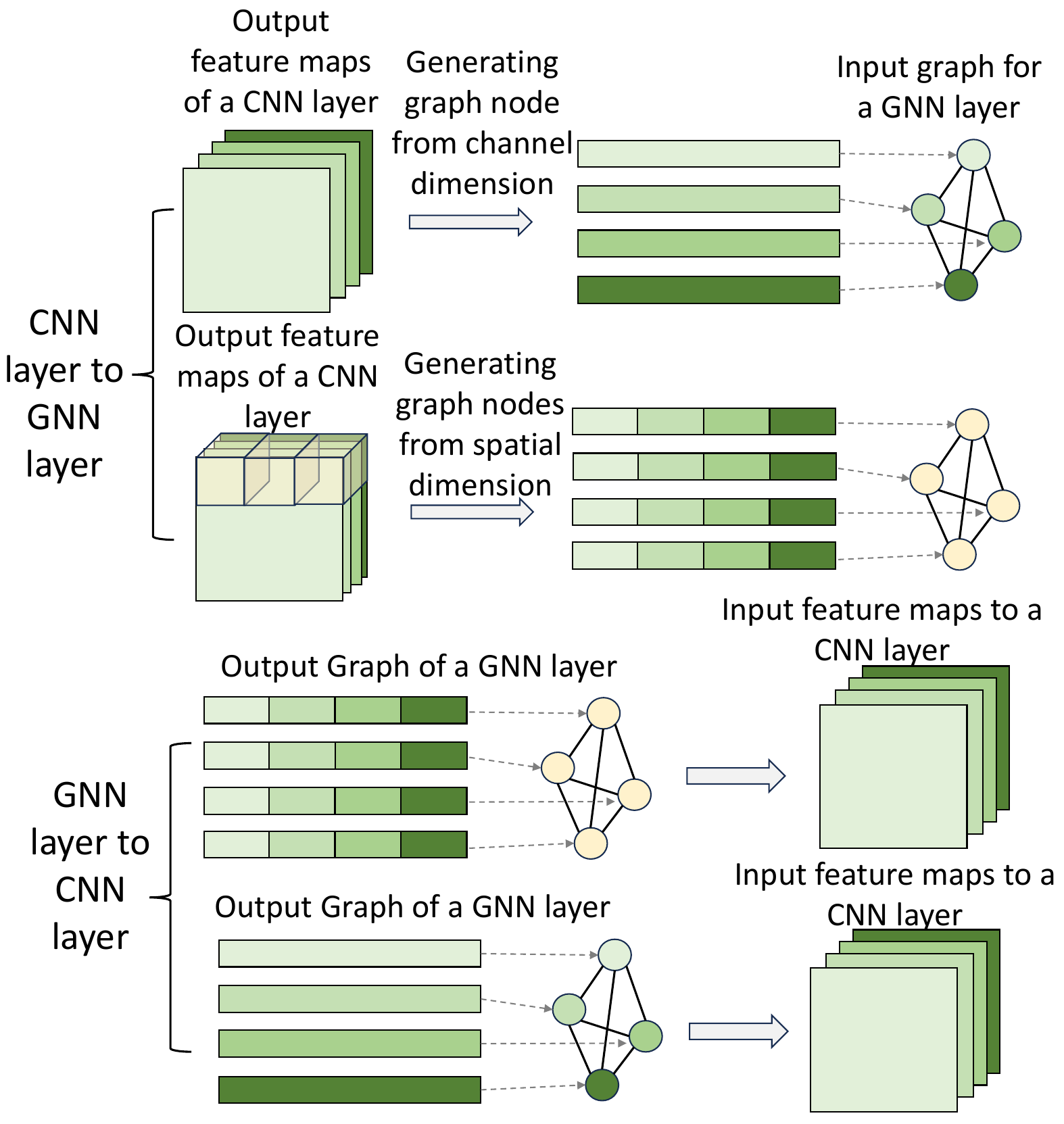}
     \caption{Data Manipulation between CNN and GNN layers}
     \vspace{-0.4cm}
     \label{fig:data-mani}
\end{figure}

CNN layer and GNN layer can have very different data layouts. For example, the output data layout of a CNN layer may not be compatible with the input data layout requirement of a GNN layer, and vice versa. The input parser generates the data manipulation (DM) layer between the CNN and GNN layers. In GNN-based CV tasks, the data manipulation process between the CNN and GNN layers is illustrated in Figure \ref{fig:data-mani}. For example, for the output feature maps of a CNN layer, GNN is used to perform reasoning in channel or spatial dimensions. For reasoning in channel dimension, each channel is viewed as a graph node (\emph{channel-to-node transformation}), while for reasoning in spatial dimension, each patch of pixels in spatial dimension is viewed as a graph node (\emph{patch-to-node transformation}). This data manipulation process can lead to significant overhead and requires careful compiler-hardware co-optimization. The DM layer will be optimized during the compilation process.

\subsubsection{Layer fusion for DM layer} To reduce the overhead of the DM layer, the compiler merges the DM layer with the following computation layer (Conv layer or MP layer). This overlaps the data manipulation and the computation. In our hardware design (Figure \ref{fig:hard-arch}), each PE maintains a Data Manipulation Module (DMM), which pipelines the data manipulation operation and computation.

\subsubsection{Uniform Mapping} CNN layer (Conv layer) and GNN layer (MP layer) have very different computation patterns. To leverage our flexible hardware design, the compiler performs uniform mapping for the CNN layer and GNN layer in step 2. Both CNN layer and GNN layer are mapped to matrix operations, including matrix multiplication and matrix addition. Since our hardware architecture is optimized for various matrix operations, both the CNN layer and GNN layer can be efficiently executed using our unified architecture design.

\subsubsection{Data Layout Centric  Mapping} 
\begin{figure}[ht]
     \centering
     \vspace{-0.2cm}
     \includegraphics[width=7cm]{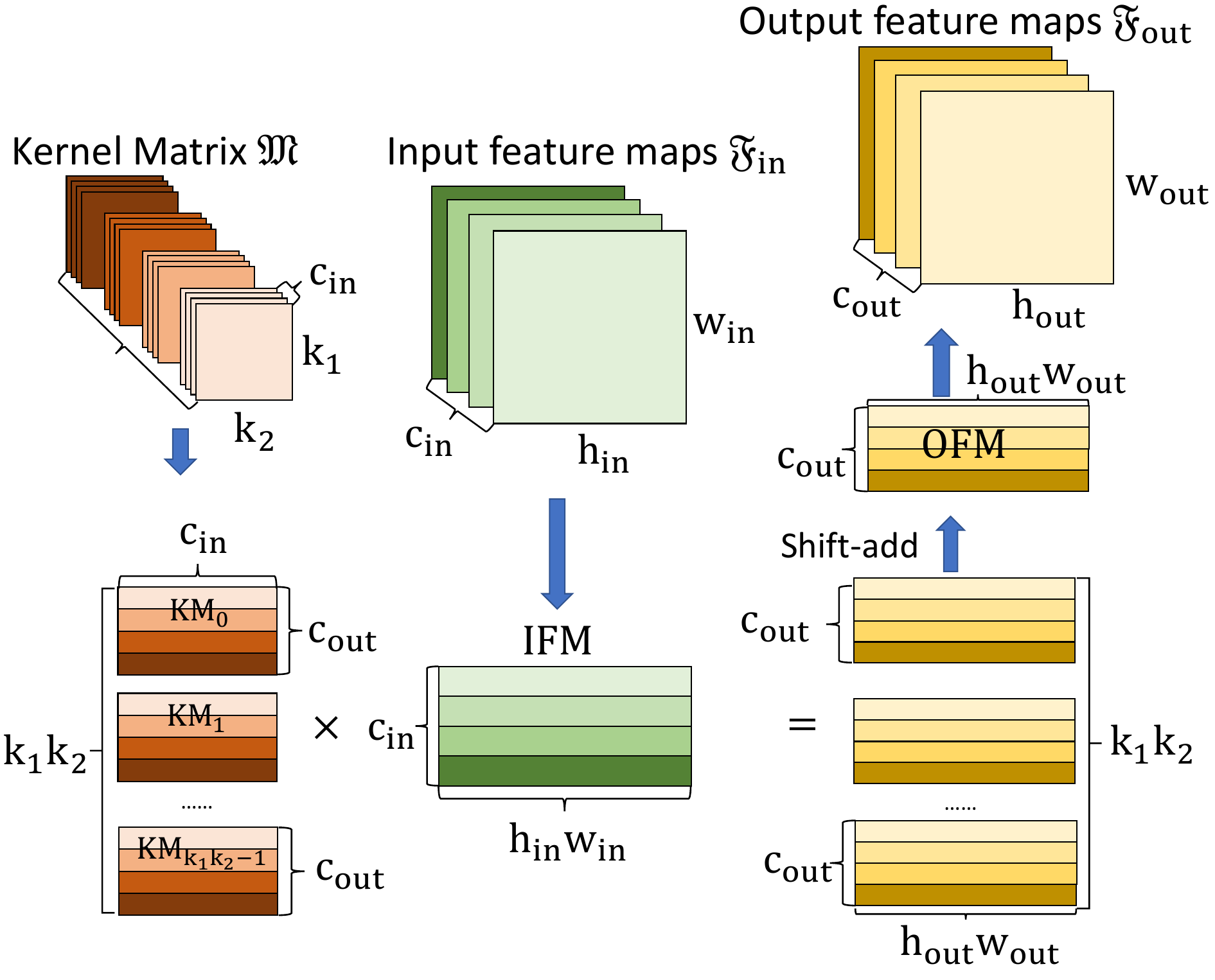}
     \vspace{-0.2cm}
     \caption{Mapping a Conv layer to matrix operations}
     \vspace{-0.2cm}
     \label{fig:kn2row-algorithm}
\end{figure}

CNN layer and GNN layer have very different data layouts. To reduce the data manipulation overhead (Figure \ref{fig:data-mani}) between two layers, we proposed to perform data layout centric mapping, which involves the mapping of Conv layers and mapping of MP layers:

\noindent \textbf{Mapping of a Conv layer}: As shown in Figure \ref{fig:kn2row-algorithm}, for a Conv layer, the convolution kernel matrix $\mathfrak{W}$ is rearranged into $k_{1}\times k_{2}$ submatrices, denoted as ${\textstyle \{\mathbf{KM}_{i}: 0 \leqslant i \leqslant k_{1}k_{2}-1\}}$, where each $\mathbf{KM}_{i}$ has dimensions $c_{\text{in}}\times c_{\text{out}}$. The input feature maps $\mathfrak{F}_{\text{in}}$ are organized into a matrix denoted $\mathbf{IFM}$ of size $c_{\text{in}} \times h_{\text{in}}w_{\text{in}}$, with each input feature map represented as a row in this matrix. Each $\mathbf{KM}_{i}$ is multiplied by $\mathbf{IFM}$ to obtain $k_{1}k_{2}$ output matrices, denoted as ${\textstyle \{\mathbf{OFM}_{i}:0 \leqslant i \leqslant k_{1}k_{2}-1\}}$, each having dimensions $c_{\text{out}}\times h_{\text{out}}w_{\text{out}}$. Through shift and add (shift-add) operations, the $k_{1}k_{2}$ output matrices are merged into a single output matrix $\mathbf{OFM}$ of size $c_{\text{out}}\times h_{\text{out}}w_{\text{out}}$. $\mathbf{OFM}$ can be further reorganized back to $c_{\text{out}}$ output feature maps. Consequently, a Conv Layer is mapped to matrix multiplication and matrix addition operations.

\vspace{0.1cm}
\noindent\textbf{Data layout of Conv Layer}: The proposed mapping strategy brings several benefits: (1) The computation of a Conv Layer is mapped to the matrix operations associated with the computation primitives. (2) The reorganization of data layout for the kernel matrix ($\mathfrak{W}$) occurs at compile time, incurring a one-time cost.  (3) The data layout for both $\mathbf{IFM}$ and $\mathbf{OFM}$ remains consistent without the need for data layout transformations between consecutive Conv Layers. (4) Most importantly, the data layout of $\mathbf{IFM}$/$\mathbf{OFM}$ can simplify the data layout manipulation between CNN layer and GNN layer. For example, if the data manipulation layer performs \emph{channel-to-node transformation}, each row of $\mathbf{IFM}$/$\mathbf{OFM}$ corresponds to a channel in the feature maps of a CNN layer. $\mathbf{IFM}$/$\mathbf{OFM}$ can serves as the input feature matrix for the following MP layer. If the data manipulation layer performs \emph{patch-to-node transformation}, each column or several columns of $\mathbf{IFM}$/$\mathbf{OFM}$ corresponds to an image patch in the feature maps of a CNN layer. The following MP layer can load node features through matrix transpose, which can be efficiently executed by the Data Manipulation Module.

\noindent\textbf{Mapping of MP layer}:  An MP layer is mapped to the multiplication of graph adjacency matrix $\mathbf{A}$ and feature matrix $\mathbf{H}$. This matrix multiplication will be mapped to either dense computation primitive (DDMM) or sparse computation primitive (SpDMM), which will be introduced later (Section \ref{subsubsec:Sparsity-aware}) in detail. To reduce the overhead of data manipulation from MP layer to Conv layer (Figure \ref{fig:data-mani}), the compiler utilizes the Buffer-to-pipeline (B2P) routing network for \emph{channel shuffling} in DDMM and SpDMM. Because for the data from a GNN layer to a CNN layer (Figure \ref{fig:data-mani}), each node feature vector or a piece of node feature vector needs to be routed to the corresponding channel of the feature maps of a CNN layer. During compilation, the compiler assigns a channel index for each node feature vector or a piece of feature vector. During hardware execution, when performing DDMM or SpDMM, the B2P routing network routes the feature vector to the corresponding channel stored in the result buffer. Through this on-the-fly channel shuffling, we eliminate the overhead of data manipulation from the GNN layer to the CNN layer.

\subsubsection{Sparsity-aware primitive mapping} \label{subsubsec:Sparsity-aware} In step 2, both CNN layers and GNN layers are mapped to matrix operations. Nevertheless, the weight matrix of a CNN/GNN layer or the adjacency matrix of a GNN layer can have different data sparsity. To exploit the data sparsity, the compiler performs sparsity-aware primitive mapping in step 4. In Step 4, for each matrix multiplication operation, the compiler maps it to dense computation primitive (DDMM) and sparse computation primitive (SpDMM) based on the data sparsity and performance models of computation primitives (Section \ref{subsec:comp-primitives}).

\section{Implementation Details}
\label{sec:implementation}

\begin{figure}[ht]
     \centering
     \includegraphics[width=7cm]{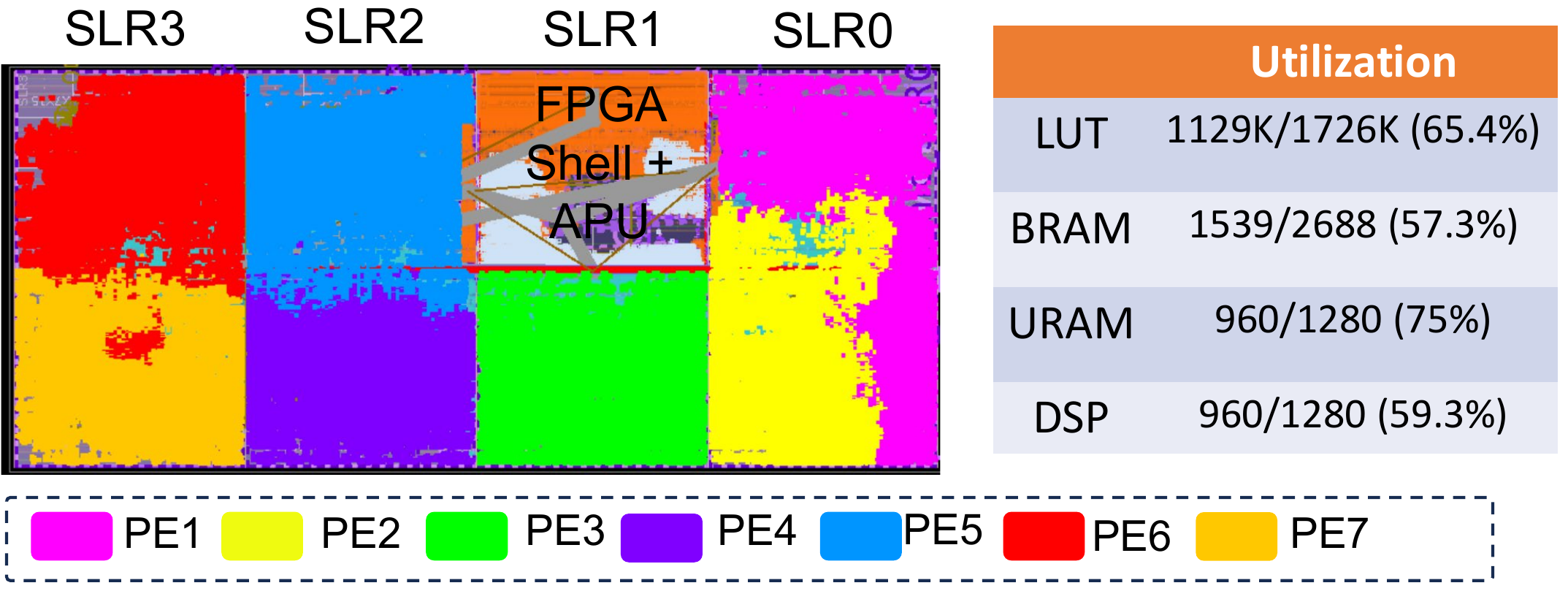}
     \caption{Device map on Alveo U250 FPGA}
     \vspace{-0.5cm}
     \label{fig:Device-map}
\end{figure}

\emph{Hardware}: We implement the accelerator and the APU (Figure \ref{fig:GCV-Turbo-Overview}) on an Alveo U250 FPGA \cite{ref-alvelu250}. We empirically set $p_{\text{ca}}=16$ for each PE and use the half-precision floating-point data format (fp16).
The Alveo U250 board consists of four Super Logic Regions (SLRs). Each SLR can be deployed with 2 PEs, except for SLR1, where half of it is occupied by FPGA shell and APU.
We utilize Verilog HDL for developing the PE, and use MicroBlaze \cite{Microblaze-link} IP core from AMD Xilinx for implementing the APU. FPGA synthesis and place-route are carried out using Vivado 2022.2. The generated device map and resource utilization are reported in Figure \ref{fig:Device-map}. We also perform frequency optimization following the methodology in Xilinx DPU \cite{dpu-clock} to set the frequency of the computation units ($f_{\text{cu}}=600$ MHz) to double that of the data buffers ($f_{\text{buffer}}=300$ MHz), enhancing the peak performance of the accelerator. 

\noindent \textbf{Impact of resource sharing}: As discussed in Section \ref{sec:Hardware-arch}, different computational primitives share the same set of computation units, on-chip buffers, and routing networks. The wires of different primitives and multiplexers for selecting data paths incur extra area costs. In each PE, these wires and multiplexers consume 37K LUTs (Figure \ref{fig:Device-map}), taking 31\% LUTs consumption of a PE (A PE consumes 118K LUTs). Through resource sharing, our PE design only costs extra 31\% LUTs for supporting various computation primitives.

\emph{Compiler}: We develop the compiler using Python built upon TVM infrastructure \cite{chen2018tvm}. Based on it, we develop our own intermediate representation, compilation workflow, and compiler optimizations. The compiler takes the computation graph generated by PyTorch \cite{paszke2019pytorch} and the metadata of input data as input. We develop customized intermediate representation (IR) as TVM prime functions. The five-step compilation is implemented as IR transformation passes to process the generated IR step by step. The output of the compiler is a sequence of instructions that is stored in a binary file.

\section{Experimental Results}
\label{sec:experiment}

\emph{Overview}: We conduct experiments to demonstrate two key aspects:
(1) \emph{Scope}: GCV-Turbo's versatility to handle a wide range of GNN-based CV tasks as well as traditional CNNs and GNNs;
(2) \emph{Performance}: GCV-Turbo's ability to achieve high performance, especially for the end-to-end acceleration of GNN-based CV tasks.
A comparison of scope and performance are summarized in Table \ref{tab:accelerator-scope} and Table \ref{tab:accelerator-performance}, respectively.
Table \ref{tab:accelerator-scope} clearly illustrates that unlike existing CNN DSAs \cite{xilinxdpu, 9065523, jouppi2018motivation, abdelfattah2018dla, yu2019opu, chen2014dadiannao, 7480791} and GNN accelerators \cite{yan2020hygcn, geng2020awb, zhang2021boostgcn, sarkar2023flowgnn, liang2020deepburning, zhang2020hardware, geng2021gcn}, which target only one specific scope (either CNNs or GNNs), GCV-Turbo can handle all three scopes -- CNNs, GNNs, as well as GNN-based CV.
We note that current state-of-the-art implementations of GNN-based CV tasks run these ML models on CPUs or GPUs \cite{implementation-b1, implementation-b2, implementation-b3, implementation-b4, implementation-b5} (given the limited scope of CNN DSAs and GNN accelerators). Thus, a natural performance comparison of GCV-Turbo is with standalone CPUs or GPUs for such tasks. Table \ref{tab:accelerator-performance} shows a comprehensive comparison of GCV-Turbo versus all alternative baselines - standalone CPU, GPU as well as all the DSAs. Note that GCV-Turbo not only offers comparable performance with CNN DSAs and GNN accelerators  within their specialized scopes, it also outperforms CPU and GPU platforms in all three scopes.
\begin{table}[!ht]
\centering
\vspace{-0.2cm}
\caption{Scope of various  accelerators}
\vspace{-0.1cm}
\begin{adjustbox}{max width=0.48\textwidth}
\begin{tabular}{c|ccc|c}
\toprule
\backslashbox{\textbf{Accelerator}}{\textbf{Scope (Models)}}                 & \begin{tabular}[|c|]{@{}c@{}}  Scope 1 \\ \textbf{(CNNs)} \end{tabular}  & \begin{tabular}[|c|]{@{}c@{}}  Scope 2 \\ \textbf{(GNNs)}  \end{tabular}  & \begin{tabular}[|c|]{@{}c@{}}  Scope 3 \\ \textbf{(GNN-based CV)} \end{tabular}  &  \begin{tabular}[|c|]{@{}c@{}}  \textbf{Performance} 
 \\ \textbf{Comparison}  \end{tabular}   \\ 
\midrule 
 CPU and GPU & \cmark & \cmark &  \cmark  & See Section \ref{subsec:Comparison-State-of-the-art} \\
 CNN DSAs \cite{xilinxdpu, 9065523, jouppi2018motivation, abdelfattah2018dla, yu2019opu, chen2014dadiannao, 7480791} &  \cmark &  \xmark &  \xmark & See Section \ref{subsubsec:cmp-cnn-dsas}\\
 GNN Accelerators \cite{yan2020hygcn, geng2020awb, zhang2021boostgcn, sarkar2023flowgnn, liang2020deepburning, zhang2020hardware, geng2021gcn}  &  \xmark & \cmark & \xmark & See Section \ref{subsubsec:cmp-gnn-dsas}  \\
\rowcolor{L2}  GCV-Turbo  &  \cmark & \cmark & \cmark  \\
\bottomrule
\end{tabular}
\end{adjustbox}
\label{tab:accelerator-scope}
\end{table}

\begin{table}[!ht]
\centering
\caption{Average speedup achieved by GCV-Turbo over various baselines within their specialized scopes. Each entry represents the performance of GCV-Turbo divided by the performance of the respective baseline. “Not supported" means that the scope is not supported by the baseline. }
\vspace{-0.2cm}
\begin{adjustbox}{max width=0.48\textwidth}
\begin{tabular}{c|ccc}
\toprule
\backslashbox{\textbf{Baseline}}{\textbf{Scope (Models)}} & \begin{tabular}[|c|]{@{}c@{}}  \textbf{CNNs} \end{tabular}  & \begin{tabular}[|c|]{@{}c@{}}  \textbf{GNNs}  \end{tabular}  & \begin{tabular}[|c|]{@{}c@{}}  \textbf{GNN-based CV tasks} \end{tabular}    \\ 
\midrule 
\rowcolor{L1} CPU (GPU) & $418.8\times$ ($1.8\times$)  & $499.5\times$ ($3.2\times$) &  $68.4\times$ ($4.1\times$)\\
\rowcolor{L2} CNN DSAs &  $0.88 - 0.93\times$ &  Not supported  &  Not supported \\
\rowcolor{L1} GNN Accelerators &  Not supported &  $1.03\times-1.25\times$ &  Not supported \\
\bottomrule
\end{tabular}
\end{adjustbox}
\label{tab:accelerator-performance}
\end{table}

The rest of this section is organized as follows: (1) Section \ref{subsec:benchmark-baseline} introduces the benchmarks, baselines, metrics, and datasets.
(2) Section \ref{subsec:Comparison-State-of-the-art} presents the comparison results with state-of-the-art CPU and GPU on six GNN-based CV tasks, and standalone CNNs and GNNs.
(3) Section \ref{subsec:impact-optimization} shows the impact of compiler optimizations.
(4) Section \ref{subsec:Comparison-with-Accelerators} compares GCV-Turbo's performance with that of state-of-the-art CNN and GNN accelerators, within their respective scopes.

\subsection{Benchmarks, Baselines, and Metrics}
\label{subsec:benchmark-baseline}

\noindent \textbf{Benchmarks}: We collect benchmarks from three scopes, including (1) \textbf{scope 3 (GNN-based CV)}: representative GNN-based CV tasks, as elaborated in Table \ref{tab:benchmark-details}, which cover diverse data modalities and model types. (2) \textbf{scope 1 (CNNs)}: popular CNN models for CV tasks, including \verb|c1|: AlexNet, \verb|c2|: ResNet-50 \cite{he2016deep}, \verb|c3|: ResNet-101 \cite{he2016deep}, \verb|c4|: VGG16 \cite{simonyan2014very}, and \verb|c5|: VGG19 \cite{simonyan2014very}; (3)  \textbf{scope 2 (GNNs)}: widely used GNN models (\verb|g1|: GCN \cite{kipf2016semi},  \verb|g2|: GraphSAGE \cite{hamilton2017inductive}, \verb|g3|: GAT \cite{velivckovic2017graph});  

\begin{table}[ht]
\centering
\caption{Details of evaluated GNN-based CV tasks}
\vspace{-0.1cm}
\begin{adjustbox}{max width=0.48\textwidth}
\begin{tabular}{ccccc}
\toprule
\textbf{Notation} & \textbf{Task} & \textbf{Input Modality} & \textbf{Model Type}  &  \textbf{Dataset}\\
\midrule
\midrule
\verb|b1| \cite{garcia2018few} & Few-shot image classification & image & CNN + GNN & Omniglot \cite{lake2019omniglot} \\ \midrule
\verb|b2| \cite{chen2019multi} & Multi-label image classification & image &  CNN + GNN &  MS-COCO \cite{lin2014microsoft} \\ \midrule
\verb|b3| \cite{zhang2019dual} & Image segmentation &  image &  CNN + GNN  & Cityscapes \cite{cordts2016cityscapes}\\ \midrule
\verb|b4| \cite{yan2018spatial} & Skeleton-based action recognition  & human skeleton &  CNN + GNN & NTU RGB+D \cite{liu2020ntu} \\ \midrule
\verb|b5| \cite{zhang2023graph} & SAR automatic target classification & radar signal & CNN + GNN &  MSTAR \cite{mstar} \\ \midrule
\verb|b6| \cite{qi2017pointnet}& Point cloud classification & point cloud &  GNN & ModelNet40 \cite{wu20153d} \\ \bottomrule
\end{tabular}
\end{adjustbox}
\label{tab:benchmark-details}
\end{table}

\begin{table}[]
\centering
\caption{Statistics of the graphs in GNN-based CV tasks}
\vspace{-0.1cm}
\begin{adjustbox}{max width=0.48\textwidth}
\begin{tabular}{cccc|cccc}
\toprule
\textbf{Model} & \textbf{\# of vertices} & \textbf{\# of edges} & \textbf{Feature length} & \textbf{Model} & \textbf{\# of vertices} & \textbf{\# of edges}
 & \textbf{Feature length}\\
\midrule
\verb|b1| & 25-100 & 300-5000 & 300-400  & \verb|b4| & 25 &75-125  & 9600-19200  \\
\verb|b2| & 80 & 6400 &  300-2048 & \verb|b5| & 16384 & 131072 &  48 \\
\verb|b3| & 100-300 & 10000-30000 & 561-33153 & \verb|b6| & 1024 &  10000-30000 &   64-1024  \\
 \bottomrule
\end{tabular}
\end{adjustbox}
\vspace{-0.2cm}
\end{table}

\vspace{0.05cm}
\noindent \textbf{Baselines}: We compare the performance with the implementations on CPU and GPU as shown in Table \ref{tab:platform-specifications}.

\begin{table}[!ht]
\centering
\caption{Specifications of platforms }
\vspace{-0.2cm}
\begin{threeparttable}
\begin{adjustbox}{max width=0.4\textwidth}
\begin{tabular}{c|ccc}
 \toprule
\textbf{Platforms} & \textbf{CPU} & \textbf{GPU}  &  \textbf{GCV-Turbo}  \\ 
\midrule 
\rowcolor{L1} Platform  & AMD Ryzen 3990x & Nvidia RTX A5000  & Alveo U250  \\
\rowcolor{L2} {Platform Technology}  & TSMC 7 nm   & Samsung 8 nm &  TSMC 16 nm\\ 
\rowcolor{L1} {Frequency} & 2.90 GHz  & 1170 MHz &  600/300 MHz \\ 
\rowcolor{L2} {Peak Performance}& 3.7 TFLOPS & 27.7 TFLOPS &  1.08 TFLOPS  \\ 
\rowcolor{L1} {On-chip Memory}& 256 MB L3 cache & 6 MB L2 cache & 45 MB   \\
\rowcolor{L2}{Memory Bandwidth}& 107 GB/s & 768 GB/s  & 77 GB/s \\ \bottomrule
\end{tabular}
\end{adjustbox}
\end{threeparttable}
\label{tab:platform-specifications}
\vspace{-0.3cm}
\end{table}

\noindent \textbf{Performance Metrics}: We consider two performance metrics (1) \emph{batch-size-one latency}: this measures the accelerator's latency when the batch size is equal to one. In applications like autonomous driving \cite{roszyk2022adopting}, low latency is critical for ensuring safety; (2) \emph{throughput}: when comparing with state-of-the-art CNN accelerators for standalone CNNs (Section \ref{tab:cnn-results}), we use throughput as the performance metric. CNN accelerator performance is typically reported in  throughput \cite{xilinxdpu, yu2019opu}.

\subsection{Comparison with CPU and GPU Implementations}
\label{subsec:Comparison-State-of-the-art}

In this section, we provide a comprehensive comparison between GCV-Turbo and CPU/GPU across the three scopes (Table \ref{tab:accelerator-scope}). The summarized results can be found in Table \ref{tab:accelerator-performance}.

\subsubsection{Evaluation on Scope 3 (GNN-based CV tasks)} 
\label{subsubsec:cpu-gpu-gnn-based-cv}

\begin{figure}[ht]
     \centering
     \vspace{-0.3cm}
     \includegraphics[width=8.5cm]{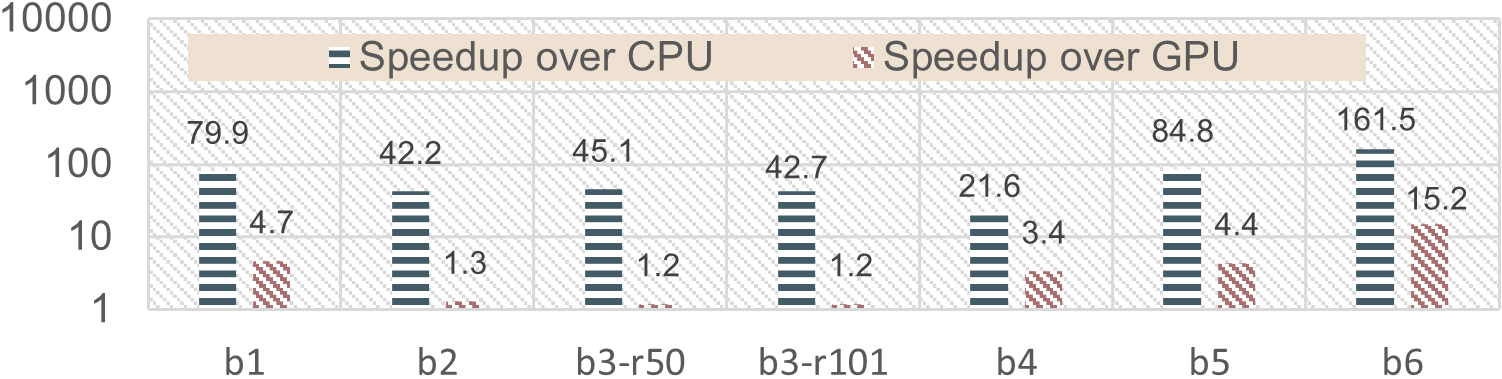}
     \vspace{-0.1cm}
     \caption{Speedup (latency reduction) over CPU and GPU on GNN-based CV tasks}
      \vspace{-0.4cm}
     \label{fig:cmp-sota}
\end{figure}

Figure \ref{fig:cmp-sota} displays the comparison results with CPU and GPU performance on six representative GNN-based CV tasks. The CPU and GPU implementations of these six tasks are from the well-optimized open-source implementations \cite{implementation-b1, implementation-b2, implementation-b3, implementation-b4, implementation-b5}, which utilize the optimized CUDA library for CNN and GNN layers.
Note that \verb|b3| employs two different CNN models, ResNet-50 and ResNet-101, in combination with their proposed GNN layers, resulting in two combinations denoted as \verb|b3-r50| and \verb|b3-r101|, respectively.
On average, GCV-Turbo achieves $68.4\times$ and $4.1\times$ \emph{latency} reduction compared with CPU and GPU, respectively. This speedup is attributed to two factors:
(1) The proposed accelerator utilizes unified architecture to accelerate both CNN and GNN layers, improving resource utilization.
(2) Our compiler optimizations hide and eliminate the overhead of data layouts transformation  between 
 CNN and GNN layers.
As illustrated in Figure \ref{fig:cmp-sota}, GCV-Turbo achieves a higher speedup on \verb|b1| and \verb|b4-6|, due to
(1) As shown in breakdown analysis below, GNN layers constitute a larger portion of the workload in \verb|b1| and \verb|b4-6|. GCV-Turbo can achieve a higher speedup for the GNN layers due to its optimized architecture for irregular computation in GNN.
(2) As shown in Table \ref{tab:model-size}, the model \verb|b1| and \verb|b4-6| can fit in the on-chip memory of our accelerator. Due to the customized on-chip memory organization, the Computation Array can access the parameters of the model in one clock cycle. In contrast, GPUs have complex cache hierarchy and small L1 cache (128 KB per SM); Accessing the parameters requires navigating through a complex cache hierarchy, resulting in higher latency.

\vspace{0.1cm}
\noindent \textbf{Discussion on the throughput of GPU}: While GCV-Turbo achieves lower latency when batch size is 1, GPU can achieve higher throughput by increasing the batch size (e.g., 8/16/32) as GPU has higher peak performance and memory bandwidth. Nevertheless, this work targets latency-sensitive applications (e.g., autonomous driving). For higher throughput, it requires FPGA vendors to develop more powerful FPGA boards with more hardware resources.

\begin{table}[ht]
\centering
\caption{Model size in GNN-based CV tasks}
\vspace{-0.1cm}
\begin{adjustbox}{max width=0.42\textwidth}
\begin{tabular}{|c|c|c|c|c|c|c|c|}
\hline
\textbf{Task} & \verb|b1| & \verb|b2| & \verb|b3-r50| & \verb|b3-r101| & \verb|b4| & \verb|b5| & \verb|b6| \\ \hline
\textbf{Model size} (MB) & 9.6 &  115 & 66 & 114 & 5.2 & 0.76 & 1.67 \\ \hline 
\end{tabular}
\end{adjustbox}
\label{tab:model-size}
 \vspace{-0.2cm}
\end{table}

 \begin{figure}[ht]
     \centering
     \includegraphics[width=8.5cm]{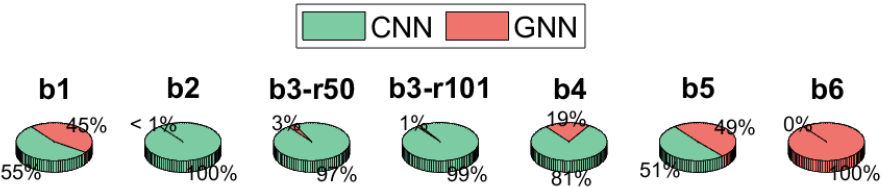}
     \caption{Proportion of hardware execution latency of various portions (CNN portion and GNN portion) on GCV-Turbo.}
     \label{fig:pie-fpga}
\end{figure}

\emph{Breakdown Analysis}: We conduct a breakdown analysis to understand the speedup of GCV-Turbo compared with the GPU on \verb|b1-6|. The results, depicted in Figure \ref{fig:pie-gpu} and \ref{fig:pie-fpga}, demonstrate that different GNN-based CV tasks consist of varying proportions of CNN and GNN layers, and data layout transformation. 
Table \ref{tab:breakdown-analysis-results} presents the breakdown analysis of speedup over the baseline GPU. GCV-Turbo achieves a speedup of $1.2-2.4\times$ on the CNN portion and $1.3-15.2\times$ on the GNN portion for various GNN-based CV tasks. Moreover, through our compiler optimizations, the overhead of data layout transformation is completely reduced and hided, leading to higher latency reduction.

\begin{table}[!ht]
\centering
\caption{Speedup (batch-size-one latency) of GCV-Turbo over GPU on various portions of the GNN-based CV tasks. For layout transformation, the speedup is $\infty$ because GCV-Turbo completely eliminates its overhead.}
\vspace{-0.1cm}
\begin{threeparttable}
\begin{adjustbox}{max width=0.45\textwidth}
\begin{tabular}{|c|c|c|c|c|c|c|c|}
\hline
  &\verb|b1| & \verb|b2| & \verb|b3-r50| & \verb|b3-r101| & $\verb|b4|$ & \verb|b5| & \verb|b6|\\ 
\hline
\textbf{CNN layers} & $2.4\times$ & $1.2\times$ & $1.2\times$ & $1.2\times$ & $1.8\times$  & $2.3\times$ &  N/A  \\ \hline
\textbf{GNN layers} & $7.6\times$ & $6.8\times$ & $1.3\times$ & $1.3\times$ & $8.4\times$ & $6.5\times$ & $15.2\times$ \\ \hline
\textbf{Layout transformation} & $\infty$ & $\infty$ & $\infty$ & $\infty$ & $\infty$ & $\infty$ & $0$ \\ \hline
\textbf{Total} & $5.1\times$ & $1.3\times$ & $1.2\times$ & $1.2\times$ & $3.6\times$ & $4.6\times$ & $15.2\times$ \\
\hline
\end{tabular}
\end{adjustbox}
\end{threeparttable}
\label{tab:breakdown-analysis-results}
\vspace{-0.2cm}
\end{table}

\subsubsection{Evaluation on Scope 1 (CNNs)}
\label{subsubsec:cpu-gpu-cnn}

Table \ref{tab:cmp-sota-cnn} illustrates the comparison between GCV-Turbo and highly-optimized CPU and GPU implementations \cite{implementation-cnn} across various widely used CNNs. On average, GCV-Turbo achieves $418.8\times$ ($1.8\times$) latency reduction compared with CPU (GPU) implementations. 

\begin{table}[ht]
\centering
\caption{Speedup (batch-size-one latency) over CPU and GPU on various CNNs}
\vspace{-0.1cm}
\begin{adjustbox}{max width=0.49\textwidth}
\begin{tabular}{|c|c|c|c|c|c|}
\hline
\textbf{Model} &  \verb|c1|: AlexNet &  \verb|c2|: ResNet50 & \verb|c3|: ResNet101 & \verb|c4|: VGG16 & \verb|c5|: VGG19 \\ \hline
\textbf{Speedup over CPU} & $182\times$ &  $43\times$ & $42\times$ & $971\times$ & $855\times$ \\ \hline
\textbf{Speedup over GPU} & $3.9\times$ &  $1.2\times$ & $1.2\times$ & $1.4\times$ & $1.5\times$ \\ \hline 
\end{tabular}
\end{adjustbox}
\label{tab:cmp-sota-cnn}
\vspace{-0.3cm}
\end{table}

\subsubsection{Evaluation on Scope 2 (GNNs)}
\label{subsubsec:cpu-gpu-gnn}

We evaluate GCV-Turbo using various GNN models and graph datasets. Table \ref{tab:cmp-gnn} displays the speedup achieved by GCV-Turbo over the CPU and GPU platforms. The implementation on CPU and GPU utilizes the state-of-the-art GNN library, PyTorch Geometric \cite{fey2019fast}. On average, GCV-Turbo achieves a speedup of $499.5\times$ compared with CPU and $3.2\times$ compared with GPU.

\begin{table}[!ht]
\centering
\caption{Speedup over CPU/GPU across various GNNs and graph datasets. [] denotes the speedup of GCV-Turbo over CPU, while () denotes the speedup of GCV-Turbo over GPU.}
\vspace{-0.1cm}
\begin{threeparttable}
\begin{adjustbox}{max width=0.48\textwidth}
\begin{tabular}{|c|c|c|c|c|c}
\hline
  & \textbf{Cora} \cite{kipf2016semi} & \textbf{CiteSeer} \cite{kipf2016semi} &  \textbf{PubMed} \cite{kipf2016semi} & \textbf{Flickr} \cite{hamilton2017inductive}\\ 
\hline
$\verb|g1|$: \textbf{GCN}  & 
    \begin{tabular}[|c|]{@{}c@{}} [$76.2\times$]  ($6.7\times$)  \end{tabular} &
    \begin{tabular}[|c|]{@{}c@{}} [$28.8\times$]   ($2.7\times$)  \end{tabular} &  
    \begin{tabular}[|c|]{@{}c@{}} [$1009\times$]   ($2.4\times$)  \end{tabular} & 
    \begin{tabular}[|c|]{@{}c@{}} [$312\times$]    ($2.4\times$)  \end{tabular} \\ \hline
$\verb|g2|$: \textbf{SAGE}  & 
    \begin{tabular}[|c|]{@{}c@{}} [$131.4\times$]  ($2.5\times$)  \end{tabular} &
    \begin{tabular}[|c|]{@{}c@{}} [$119.7\times$]  ($1.9\times$)  \end{tabular} &
    \begin{tabular}[|c|]{@{}c@{}} [$178.9\times$]  ($2.1\times$)  \end{tabular} &
    \begin{tabular}[|c|]{@{}c@{}} [$421.9\times$]  ($3.6\times$)  \end{tabular} \\ \hline
$\verb|g3|$: \textbf{GAT}  &
    \begin{tabular}[|c|]{@{}c@{}} [$2250\times$]   ($6.8\times$)  \end{tabular} &
    \begin{tabular}[|c|]{@{}c@{}} [$1016\times$]   ($2.9\times$)  \end{tabular} &  
    \begin{tabular}[|c|]{@{}c@{}} [$178.9\times$]  ($2.1\times$)  \end{tabular} &
    \begin{tabular}[|c|]{@{}c@{}} [$278.8\times$] ($2.0\times$)  \end{tabular} 
    \\
\hline
\end{tabular}
\end{adjustbox}
\end{threeparttable}
\label{tab:cmp-gnn}
\end{table}

\subsection{Impact of Compiler Optimizations}
\label{subsec:impact-optimization}
We evaluate the impact of two compiler optimizations:

\noindent \textbf{Layer fusion}: 
Layer fusion yields a speedup ranging from $11.8\%$ to $48.9\%$ across the six GNN-based CV tasks. This speedup can be attributed to layer fusion's capacity to enhance task-level parallelism, reduce external memory traffic, and decrease the overall computational complexity.


\noindent \textbf{Sparsity-aware mapping}: As the weight matrices of the CNN portions in \verb|b1|-\verb|b6| remain unpruned, sparsity-aware mapping does not accelerate the CNN portion. As a result, our speedup measurements exclusively focus on the GNN portion within \verb|b1|-\verb|b6|. The sparsity-aware mapping results in speedup percentages of $5.2\%$, $330\%$, $356\%$, $356\%$, $2.3\%$, $2.3\%$, $20.5\%$, and $0\%$ for the GNN portions within \verb|b1| to \verb|b6|, respectively. The GNN within \verb|b6| does not experience any speedup because, in \verb|b6|, the GNN consists of Linear layers, activation layers, and batch normalization layers. The weight matrices within the Linear layers of \verb|b6| do not have data sparsity.

\subsection{Comparison with State-of-the-art Accelerators}
\label{subsec:Comparison-with-Accelerators}

We compare the performance of GCV-Turbo with CNN DSAs \cite{xilinxdpu, yu2019opu} on CNN models in Section \ref{subsubsec:cmp-cnn-dsas} and with GNN accelerators \cite{zhang2021boostgcn, ref-graphagile} on GNN models in Section \ref{subsubsec:cmp-gnn-dsas}.
Different accelerators are implemented on different hardware platforms and use different amount of hardware resources. For a fair comparison, we normalize the performance (latency/throughput) by their respective peak performance (FLOPs). For example, normalized throughput is calculated by:
$
    {\textstyle\scalebox{1}{\text{Normalized Throughput of [X]}} = \frac{\scalebox{0.7}{\text{Throughput of [X]}}}{\scalebox{0.7}{\text{Peak performance of [X]}}}}
    \label{eq:normalized-speedup}
$
where $\text{[X]}$ can be AMD DPU \cite{xilinxdpu}, OPU \cite{yu2019opu}, BoostGCN \cite{zhang2021boostgcn}, GraphAGILE \cite{ref-graphagile}, or GCV-Turbo.

\begin{table}[!ht]
\centering
\vspace{-0.2cm}
\caption{Specifications of CNN/GNN accelerators}
\vspace{-0.1cm}
\begin{threeparttable}
\begin{adjustbox}{max width=0.48\textwidth}
\begin{tabular}{c|cc|cc}
 \toprule
& \multicolumn{2}{c|}{CNN DSAs } & \multicolumn{2}{c}{GNN Accelerators }  \\ 
\textbf{Platforms} & AMD DPU \cite{xilinxdpu} & OPU1024 \cite{yu2019opu} &   BoostGCN \cite{zhang2021boostgcn} &  GraphAGILE \cite{ref-graphagile} \\ 
\midrule \midrule 
\rowcolor{L1} Platform  & ZCU102 &  Xilinx XC7K325T  & Stratix10 GX &  Alveo U250  \\
\rowcolor{L2} {Platform Technology}  & N/A   & 28 nm  & Intel 14 nm   & TSMC 16 nm \\ 
\rowcolor{L1} {Peak Performance}& 1.15 TFLOPS & 0.2 TFLOPS & 0.64 TFLOPS & 0.64 TFLOPS \\ 
\rowcolor{L2} {On-chip Memory}& 32.1 MB & 2 MB  & 45 MB  & 45 MB   \\
\rowcolor{L1} {Memory Bandwidth}& 19.2 GB/s & 12.8 GB/s  & 77 GB/s  & 77 GB/s\\ \bottomrule
\end{tabular}
\end{adjustbox}
\end{threeparttable}
\label{tab:platform-CNN-DSA}
\end{table}

\subsubsection{Comparison with CNN domain-specific accelerators (DSAs)} \label{subsubsec:cmp-cnn-dsas} We compare GCV-Turbo's performance with state-of-the-art FPGA-based CNN DSAs,  AMD DPU \cite{xilinxdpu} and OPU \cite{yu2019opu} (Table \ref{tab:platform-CNN-DSA}),   on throughput (Table \ref{tab:cnn-results}). GCV-Turbo's throughput is computed as $\frac{1}{\text{latency}}$.
GCV-Turbo achieves a normalized throughput of $0.88\times$ and $0.93\times$ compared with OPU and DPU on \verb|c1-c5|. These results demonstrate GCV-Turbo's competitive throughput in various CNN models, despite slight lower performance. These throughput differences are due to two design trade-offs:
GCV-Turbo's versatility, supporting both CNNs and GNNs, sacrifices some CNN-specific architectural optimizations. For example, OPU's multi-level parallelism is fine-tuned for CNN convolution operations, whereas GCV-Turbo's architecture is more generalized.
GCV-Turbo's compilation flow optimizes CNNs and GNNs holistically but cannot support certain convolution-specific optimizations. DPU, for example, selects dataflow for convolutional layers based on kernel size, which cannot be directly applied to GNN layers. Moreover, due to CNN-specific optimizations, OPU and DPU have higher efficiency for using limited DDR memory bandwidth for CNNs.
However, OPU and DPU compilers do not support GNNs, and their architectures are inefficient for irregular computations and memory access patterns of GNNs.





\begin{table}[!ht]
\centering
\caption{Comparison of inference throughput (images/second) with CNN DSAs on various CNN models }
\begin{threeparttable}
\begin{adjustbox}{max width=0.45\textwidth}
\begin{tabular}{c|ccccc|c}
 \toprule
  & \multicolumn{5}{c|}{\textbf{Throughput (unnormalized)}} &  \multirow{2}{*}{ \begin{tabular}[|c|]{@{}c@{}} Normalized average speedup  of \\ GCV-Turbo over the DSA \end{tabular} } \\
  & \begin{tabular}[|c|]{@{}c@{}} \verb|c1|  \end{tabular}  & 
  \begin{tabular}[|c|]{@{}c@{}} \verb|c2| \end{tabular}  & 
  \begin{tabular}[|c|]{@{}c@{}} \verb|c3|  \end{tabular}  & 
  \begin{tabular}[|c|]{@{}c@{}} \verb|c4| \end{tabular} & 
  \begin{tabular}[|c|]{@{}c@{}} \verb|c5| \end{tabular} &
 \\ 
\midrule \midrule 
\textbf{DPU} \cite{xilinxdpu}& N/A & $43.4$ & $38.8$ & $\textbf{274}$ & N/A  & $0.93\times$  \\ \midrule
\textbf{OPU1024} \cite{yu2019opu} & N/A & $12.2$ & $9.7$ & $54.4$ & $27$  & $0.88\times$ \\ \midrule
\textbf{GCV-Turbo} & $512.9$ & $\textbf{58.8}$ & $\textbf{46.5}$ & $254.7$ & $\textbf{127.3}$ & $1\times$  \\
\bottomrule
\end{tabular}
\end{adjustbox}
\end{threeparttable}
\label{tab:cnn-results}
\vspace{-0.1cm}
\end{table}

\subsubsection{Comparison with GNN Accelerators} \label{subsubsec:cmp-gnn-dsas}  
We compare GCV-Turbo with state-of-the-art GNN accelerators, BoostGCN \cite{zhang2021boostgcn}, GraphAGILE \cite{ref-graphagile}, and FlowGNN \cite{sarkar2023flowgnn} (see Table \ref{tab:platform-CNN-DSA}). Latency measurements follow the methodology from \cite{zhang2021boostgcn, ref-graphagile}, focusing on GCN model and various \textbf{non-CV} graph datasets (Citation networks: CO \cite{kipf2016semi}, CI \cite{kipf2016semi}, PU \cite{kipf2016semi}; Recommendation systems: FL \cite{zeng2019graphsaint}, RE \cite{hamilton2017inductive}, YE \cite{zeng2019graphsaint}, AP \cite{zeng2019graphsaint}). Table \ref{tab:speedup-over-GNN-accelerators} presents the results, where latency is normalized by peak performance of the hardware platform to obtain the speedup.
GCV-Turbo outperforms BoostGCN and GraphAGILE with speedups of $1.25\times$ and $1.03\times$, respectively. BoostGCN's inferior performance is attributed to separate sparse and dense computation hardware modules, leading to underutilization. In contrast, GCV-Turbo optimizes resource usage with a unified architecture for both sparse and dense computations in GNNs.
The slight advantage over GraphAGILE is due to GCV-Turbo's sparsity-aware mapping (Section \ref{subsubsec:Sparsity-aware}), considering data sparsity in the input graph's connectivity. GCV-Turbo maps computations to DDMM for densely connected subgraphs, unlike GraphAGILE, which neglects data sparsity in graph connectivity.
Compared with FlowGNN, GCV-Turbo performs lower because (1) FlowGNN leverages sparsity in graph feature matrices (CO, CI, PU has high data sparsity ($>90\%$) in feature matrices), while GCV-Turbo only uses sparsity in graph adjacency and weight matrices, and (2) FlowGNN generates optimized hardware implementation for different input models. However, the above two optimizations of FlowGNN are unattractive for CV tasks because (1) the sparsity of graph feature matrices is only known during hardware execution; Utilizing its sparsity needs on-the-fly sparsity profiling and data format transformation, causing extra preprocessing overhead. Moreover, the execution time varies with the sparsity of the input data. However, autonomous driving requires deterministic latency for safety. (2) An autonomous driving system will execute various models for various data modalities. Generating optimized bitstreams for each model incurs large latency for switching between the bitstreams through dynamic reconfiguration.



\begin{table}[!ht]
\centering
\caption{Comparison of hardware execution latency (ms) with state-of-the-art GNN accelerators}
\vspace{-0.1cm}
\begin{threeparttable}
\begin{adjustbox}{max width=0.48\textwidth}
\begin{tabular}{c|ccccccc|c}
 \toprule
   & \multicolumn{7}{c|}{\textbf{Latency (ms) (unnormalized)}} &  \multirow{2}{*}{ \begin{tabular}[|c|]{@{}c@{}} Normalized average speedup  of \\ GCV-Turbo over the accelerator \end{tabular} } \\
  & \textbf{CO} & \textbf{CI} & \textbf{PU} & \textbf{FL} & \textbf{RE} & \textbf{YE} & \textbf{AP}  &   \\ 
\midrule \midrule 
\textbf{BoostGCN} &  
        N/A &  
        N/A &  
        N/A &  
       $20.1$ &  
       $98.5$ &  
       $193.5$  &  
       $793.5$ & 
       $1.25\times$\\ \midrule
\textbf{GraphAGILE} &  
      $0.819$  & 
      $2.55$  &  
      $2.24$   & 
      $11.5$   &  
      $97.2$  &  
      $104.3$  &  
      $315$ &
      $1.03\times$\\ \midrule
\textbf{FlowGNN} &  
      $6.9E-3$  & 
      $8.3E-3$  &  
      $53E-3$   & 
      $N/A$   &  
      $136$  &  
      $N/A$  &  
      $N/A$ &
      $0.003\times$ (CO/CI/PU), $0.38\times$ (RE)\\ \midrule
\textbf{GCV-Turbo} &
     $0.48$   &  
     $1.47$   &  
     $1.25$    &  
     $6.09$    &  
     $72.7$   &  
     $43.5$    & 
     $196.9$  & 
     $1\times$\\
\bottomrule
\end{tabular}
\end{adjustbox}
\end{threeparttable}
\label{tab:speedup-over-GNN-accelerators}
\vspace{-0.3cm}
\end{table}

\section{Conclusion and Future Work}
\label{sec:conclusion-and-future-work}

We introduced GCV-Turbo, the first domain-specific accelerator for GNN-based CV. It bridges the gap between state-of-the-art CNN DSAs and GNN accelerators, leading to end-to-end acceleration of GNN-based CV. GCV-Turbo can also handle standalone CNNs and GNNs as effectively as specialized accelerators while optimizing the execution of CNN and GNN layers within GNN-based CV tasks.
With its unified architecture and compilation workflow, GCV-Turbo achieved average latency reduction of $68.4\times$ and $4.1\times$  compared with state-of-the-art CPU and GPU implementations.
In the future, we plan to broaden GCV-Turbo's capabilities to accommodate Vision Transformer (ViT) models. 




\section*{Acknowledgement}
This work is supported by the DEVCOM Army Research Lab (ARL) under grants W911NF2220159, and the National Science Foundation (NSF) under grants CCF-1919289. Equipment and support by AMD AECG are greatly appreciated.

\noindent\textbf{Distribution Statement A}: Approved for public release. Distribution is unlimited.

\bibliographystyle{IEEEtran}
\bibliography{reference}

\end{document}